\documentclass[a4paper,11pt]{article}
\usepackage{jcappub}

\usepackage{graphicx} 
\usepackage[dvipsnames]{xcolor}
\hypersetup{colorlinks, urlcolor=Blue, citecolor=Blue, linkcolor=Blue, pdfusetitle}
\usepackage{booktabs}
\usepackage{orcidlink}
\usepackage{multirow}
\usepackage{makecell}
\usepackage{fontawesome5}
\usepackage{soul}

\definecolor{darkorange}{RGB}{255, 140, 0} 

\renewcommand{\cellalign}{tl}

\title{Localizing binary neutron stars with the Einstein Telescope: a full Bayesian comparison of the triangular and 2L configurations}

\author[a,b]{Filippo Santoliquido\,\orcidlink{0000-0003-3752-1400}}
\author[a,b,c]{Jacopo Tissino\,\orcidlink{0000-0003-2483-6710}}
\author[a,b]{Marica Branchesi\,\orcidlink{0000-0003-1643-0526}}
\author[d]{Ulyana Dupletsa\,\orcidlink{0000-0003-2766-247X}}

\affiliation[a]{Gran Sasso Science Institute (GSSI), I-67100 L'Aquila, Italy}
\affiliation[b]{INFN, Laboratori Nazionali del Gran Sasso, I-67100 Assergi, Italy}
\affiliation[c]{Theoretisch-Physikalisches Institut, Friedrich-Schiller-Universität Jena, 07743 Jena, Germany}
\affiliation[d]{Marietta Blau Institute - Austrian Academy of Sciences, 1010 Vienna, Austria}

\emailAdd{filippo.santoliquido@gssi.it}

\abstract{The Einstein Telescope (ET), the European next-generation gravitational-wave observatory, is at a decisive stage: the choice of its detector configuration is still open, with profound implications for its scientific output. We provide a quantitative basis for this choice by comparing two candidate configurations at their proposed sites: the triangular design ($\Delta$) at the Euregio Meuse--Rhine site, and the 2L design, with interferometers in Sardinia, Italy, and Lusatia, Germany. 
We present the first realistic estimates of sky-localization uncertainties for binary neutron star (BNS) mergers, moving beyond Fisher-matrix forecasts to full Bayesian parameter estimation that captures the multimodal structure of the sky posterior. We focus mainly on the high-frequency interferometers alone, providing an expected scenario for the first years of ET operations, as the more challenging low-frequency instruments will come online later.
BNS injections span different sky positions and luminosity distances between 40 and 2000~Mpc, corresponding to signal-to-noise ratios (SNRs) from 12 to 720. Our analysis relaxes the long-wavelength approximation and includes Earth's rotation and higher-order modes, all of which improve sky localization. We find that the $\Delta$ configuration localizes low- and moderate-SNR sources in eight degenerate sky modes spread across both hemispheres and over the day and night sides of the Earth, with $90\%$ credible areas reaching $\sim 2 \times 10^4$~deg$^2$. In contrast, the 2L configuration recovers a single dominant mode that contains at least $50\%$ of the posterior probability and is one to two orders of magnitude smaller in area. This represents a substantial advantage for multimessenger follow-up campaigns. 
}

\begin{document}

\maketitle
\flushbottom

\section{Introduction}


The Einstein Telescope \citep[ET,][]{Punturo:2010zz,ET:2019dnz,Branchesi:2023mws,ET:2025xjr}, together with other ground-based next-generation gravitational-wave (GW) observatories such as Cosmic Explorer \citep[CE,][]{Reitze:2019iox,Evans:2021gyd,Evans:2023euw,Gupta:2023lga}, will detect tens of thousands of binary neutron star (BNS) mergers per year, owing to their improved sensitivity across the entire frequency band. A large fraction of these events will be accompanied by a detectable electromagnetic (EM) counterpart \citep{Ronchini:2022gwk,Banerjee:2022gkv,Loffredo:2024gmx,Colombo:2025sdm,Bisero:2025tkw,CTAOConsortium:2026rnr,Agarwal:2026wsp,Biswas:2026shz,Kaur:2024yag,Dobie:2019eeg,Steinle:2025xae,Han:2025fii,Khadkikar:2025ith}, spanning the spectrum from radio to gamma rays; as a result, ET will detect hundreds of multimessenger events per year.

The multi-messenger observation of GW170817/AT2017gfo \citep{LIGOScientific:2017vwq,LIGOScientific:2017ync,LIGOScientific:2017zic} has already demonstrated the scientific potential of these joint detections. A substantial sample of multi-messenger sources will therefore have a revolutionary impact across astrophysics, enabling us to observe the site of heavy-element nucleosynthesis \citep{Pian:2017gtc,Kasen:2017sxr,Smartt:2017fuw}, study the launch and
properties of relativistic jets \citep{Mooley:2018qfh,Ghirlanda:2018uyz,Lazzati:2017zsj,Troja:2017nqp},  probe the behavior of matter at extreme densities \citep{LIGOScientific:2018cki,Annala:2017llu,De:2018uhw}, and measure cosmological parameters \citep{LIGOScientific:2017adf}.

The prospects for detecting the EM counterpart degrade as the localization region estimated from the GW signal grows, since a larger area on the sky must be covered before a counterpart can be found and observed also with instruments operating in the ET era, such as the Vera Rubin Observatory \citep{LSST:2008ijt}, the Wide-field
Spectroscopic Telescope \citep[WST,][]{WST:2024rai}, CTAO \citep{CTAConsortium:2013ofs}, newAthena \citep{Nandra:2013jka}, and the SKA observatory \citep{Weltman:2018zrl}. It is therefore necessary to study the BNS localization performance of ET under different configurations using the most accurate data-analysis methods, i.e. a full Bayesian analysis that goes beyond the Fisher-information matrix approximation \citep{Iacovelli:2022bbs,Branchesi:2023mws,Maggiore:2024cwf,Dupletsa:2026jzs}.

Two geometries are currently under consideration for ET: a triangular configuration ($\Delta$), consisting of three nested 10-km interferometers, and a 2L design, consisting of two 15-km L-shaped detectors. Both are xylophone observatories \citep{Hild:2010id}, pairing a high-frequency (HF) instrument with a cryogenic low-frequency (LF) one. Since the LF instrument poses the more demanding technological challenges, the HF instruments are expected to start taking data first, although no operational timeline has been officially established. In this work we therefore target mainly the HF-only configuration, which represents the ET observing scenario expected for the first years of operation.

In \textit{Santoliquido et al. (2025)} \cite{Santoliquido:2025lot}, we showed that ET in its triangular configuration will localize short-duration sources, such as high-redshift binary black holes, in eight distinct and disconnected sky patches \citep{Singh:2021bwn,Singh:2020wsy}, or modes, i.e. local maxima of the posterior distribution, each containing a significant fraction of the posterior probability. In \textit{Santoliquido et al. (2026)} \cite{Santoliquido:2025aiq}, we showed that the 2L configuration instead localizes the majority of high-redshift sources in far fewer sky modes, but it can introduce multimodalities in the one-dimensional posterior of the luminosity distance for a subset of sources. In this work, we focus on the localization of BNS mergers, to assess under which conditions these multiple modes persist for long-duration signals.

Forecasting the ability of next-generation GW detectors to recover BNS parameters through full Bayesian analysis is an emerging line of research. For instance, \textit{Nitz and Dal Canton (2021)} \cite{Nitz:2021pbr} examined the prospects for pre-merger detection and localization across a broad range of detector networks, without however addressing the number of sky modes, while \textit{Puecher et al. (2023)} \cite{Puecher:2023twf} focused on the intrinsic parameters of BNS sources, comparing ET designs with different arm lengths, geometries, and alignments, but did not consider localization. More recently, \textit{Koehn et al. (2026)} \citep{Koehn:2026nin} forecast multimessenger detections for the $\Delta$ and 2L configurations, running full Bayesian and hierarchical inference on the events with a kilonova counterpart, but relied on the Fisher-information matrix approximation for sky localization. Further studies, such as \cite{Zhao:2017cbb,Nitz:2020vym,Smith:2021bqc,Vretinaris:2025wdu,Huez:2025gja,Sengupta:2026hif, OShaughnessy:2026kkf}, have explored related aspects of this problem.

Analyzing a BNS signal while accounting for all the challenges expected with next-generation GW detectors remains an open problem \citep{Couvares:2021ajn,Hu:2024mvn,Baker:2025taj,ET:2025xjr}, mainly because of the long duration and wide bandwidth of the signal. In the context of CE, \textit{Baral et al. (2023)} \cite{Baral:2023xst} and \textit{Baral et al. (2025)} \cite{Baral:2025geo} performed, for the first time, unbiased parameter estimation of high-SNR BNS signals (up to an SNR of $\sim1000$),
showing that accounting for the frequency dependence of the antenna patterns improves
the sky localization of compact binary coalescences, even with a single detector.


In this work, we perform full Bayesian parameter estimation of GW170817-like BNS signals for ET, comparing its two possible configurations, the triangular and the 2L. By varying the distance, sky position and inclination of the source, we analyze a broad range of BNS signals, from above the detection threshold (SNR of 12) to high-SNR events (SNR above 720), and quantitatively assess the sky-localization performance of the two configurations.



The manuscript is organized as follows. Section~\ref{sec:detectors} introduces the detector configurations, and Section~\ref{sec:pe} describes the parameter estimation and injection setup and the post-processing of the source posteriors. Section~\ref{sec:low_snr} presents the results for low- and moderate-SNR BNS events, for which we also vary the true sky position, and Section~\ref{sec:low_freq} extends the lowest-SNR case to the full design sensitivity. Section~\ref{sec:high_snr} focuses on a GW170817-like event, which would be detected at high SNR in the ET band. We discuss the implications of our results in Section~\ref{sec:discussion} and draw our conclusions in Section~\ref{sec:conclusion}.

\section{Methods}

\subsection{Detector configurations}
\label{sec:detectors}

We consider the $\Delta$ configuration at the Euregio Meuse--Rhine (EMR) site, and the 2L geometry with one interferometer in Sardinia (Italy) and an identical one in Lusatia (Germany), with a 45° angle between the arms of the two instruments. We adopt the sensitivity curves recently released in \textit{Dupletsa et al. (2026)} \cite{Dupletsa:2026jzs}, which incorporate a few upgrades over those used in recent ET collaboration works \citep{Branchesi:2023mws,ET:2025xjr}. These curves are not site-dependent and therefore differ only through the arm lengths. 

For most of our runs we restrict the analysis to the HF instruments. This is a realistic assumption for the first years of operation, and also a practical necessity: including the LF band extends the signal to several hours and makes the full likelihood evaluation prohibitively expensive at the SNRs considered here, as we quantify in Section~\ref{sec:low_freq}. These runs cover $f_\mathrm{min} = 6$~Hz to $f_\mathrm{max} = 4096$~Hz, using 4896-second time segments, corresponding to approximately 1 hour and 22 minutes. The single exception is the run presented in Section~\ref{sec:low_freq}, the only one performed with the full design sensitivity curve and starting at 3~Hz, which we carry out at the lowest SNR considered here, where the analysis remains computationally feasible. Coordinates, arm orientations and the adopted sensitivity curves are given in Appendix~\ref{sec:detector_appendix}.

\subsection{Parameter estimation}
\label{sec:pe}

The long duration and wide bandwidth of BNS signals in next-generation GW detectors make the antenna patterns frequency-dependent. On one side, the wavelength of the observed GWs becomes comparable to the interferometer dimensions, invalidating the long-wavelength approximation (LWA); on the other, the rotation of the Earth during data acquisition becomes relevant, since these signals can last from several minutes to hours within the sensitivity band of these detectors. Appendix~\ref{sec:fdep} details both effects and shows their impact on the amplitude and phase of a BNS signal.

We study a representative set of BNS injections using {\sc Bilby-xg} \citep{bilby_xg}, a fork released in \textit{Baral et al. (2025)} \cite{Baral:2025geo} of {\sc Bilby} \citep{Romero-Shaw:2020owr}, a Python library for Bayesian inference designed for GW data analysis. It extends {\sc Bilby} with two key features. The first is the implementation of the frequency-dependent antenna patterns described above. The second is relative binning \cite{Zackay:2018qdy,Dai:2018dca,Leslie:2021ssu,Cornish:2021lje}, a method that accelerates likelihood evaluations by computing the GW signal on a reduced frequency grid and interpolating between bins, while maintaining high accuracy; this implementation also accounts for higher-order modes (HOMs), i.e. the expansion of the GW signal beyond the quadrupole emission. Further details are given in Appendix~\ref{sec:likelihood_appendix}.

The posterior probability distribution of the source parameters $\theta$ is given by the Bayes' theorem:
\begin{equation}
    p(\theta|d) = \frac{\mathcal{L}(d|\theta)\,\pi(\theta)}{\mathcal{Z}(d)} \, ,
\end{equation}
where $\mathcal{L}(d|\theta)$ is the likelihood defined in Appendix~\ref{sec:likelihood_appendix}, $\pi(\theta)$ the prior, and $\mathcal{Z}(d)$ the evidence. We consider a set of zero-noise injections of BNS signals whose parameters are fixed to the median posterior values of GW170817 \citep{LIGOScientific:2017vwq}, with three exceptions. The sky position is either set to the precisely known localization of GW170817 or randomly sampled from the prior; the luminosity distance is varied from 40 to 2000~Mpc to span a broad range of optimal SNRs; and the inclination angle is fixed to either 20° or 60°. The two inclinations are chosen to bracket two qualitatively different multi-messenger scenarios: 20° corresponds to a GW170817-like source, mildly off-axis, for which the relativistic jet afterglow is expected to be detectable \citep{Colombo:2022zzp}, whereas 60° represents a strongly off-axis configuration in which the jet emission is largely beamed away and the electromagnetic counterpart is dominated by the sub-relativistic ejecta. Appendix~\ref{sec:injection_appendix} provides further details on the parameters used to perform the BNS injections, as well as on the adopted priors.  We sample the posterior using {\sc nessai} \citep{Williams:2021qyt,Williams:2023ppp}, a nested sampling algorithm enhanced with normalizing flows and implemented in {\sc {Bilby}}.

Since we are interested in the sky-localization performance of the different ET configurations, we neglect tidal effects, which are known to correlate mostly with the intrinsic parameters, such as masses and spins \citep{Wade:2014vqa, Dudi:2018jzn, Puecher:2023twf}. This is further proved in Appendix~\ref{sec:tides} where we analyze one source in the two configurations with \texttt{TEOBResumSPA} waveform approximant  \citep{Akcay:2018yyh,Gamba:2020ljo,Gamba:2023mww}, confirming that including tidal effects does not affect our qualitative conclusions on sky localization. 

We therefore adopt the frequency-domain approximant \texttt{IMRPhenomXPHM} \citep{Pratten:2020ceb,lalsuite}, which models the inspiral, merger and ringdown of quasi-circular black hole binaries. Beyond the dominant $(2,|2|)$ mode, it includes the $(\ell, |m|) = (2,|1|), (3,|3|), (3,|2|), (4,|4|)$ harmonics. Although
this approximant also models precession, we neglect it and consider only aligned-spin systems. This is a reasonable choice for BNSs, whose components are expected to be slowly spinning ($\chi_i \lesssim 0.05$), in line with the Galactic BNS population \citep{Burgay:2003jj,Hessels:2006ze,Lynch:2011aa} and with the absence of measurable spin in GW170817 \citep{LIGOScientific:2017vwq,LIGOScientific:2018hze}.


We also quantify the bias introduced by neglecting the frequency-dependent antenna patterns (Appendix~\ref{sec:fdep}) and the HOMs, treating each effect separately and one at a time, and only in the likelihood. That is, the signals were always injected with the full physics, while a given effect was discarded solely in the recovery.

For each BNS injection, we quantify localization performance across detector configurations using the \texttt{ligo.skymap} library \citep{Singer:2015ema}. We compute the sky area of the smallest 90\% credible regions ($\Delta \Omega_{90\%}$) from sky maps sampled on equal-area HEALPix pixels \citep{Gorski:1999rt,Gorski:2004by}. Incorporating the luminosity-distance posterior, we further compute the 90\% highest-probability-density comoving volumes ($\Delta V^c_{90\%}$) \cite{Singer:2016erz,Singer:2016eax}.

Furthermore, we quantify the degree of multimodality in the sky-localization posterior through the metric $P_{\text{main}}$, defined as the fraction of posterior probability contained in the mode centered on the true sky position. We estimate it as the percentage of posterior samples whose right ascension and declination lie within a circular region of 300~deg$^2$ centered on the injected location, with angular separations computed on the sphere. The radius of this region ($\sim 10$°) is chosen to be large enough to enclose the main mode for all the sources considered here. A perfectly unimodal posterior yields $P_{\text{main}} = 100\%$, while secondary modes displaced from the true position reduce it. This metric lets us track the transition between unimodal and multimodal localizations quantitatively.


\section{Results}


\subsection{Low- and moderate-SNR events}
\label{sec:low_snr}

\begin{figure*}[tp]
\centering
\newcommand{\rowtitle}[1]{\par\vspace{8pt}{\scriptsize\textbf{#1}}\par\vspace{-3pt}}

\vspace{-8pt}
\rowtitle{Moderate SNR, high-frequency sensitivity}
\begin{minipage}[t]{0.6\textwidth}
    \vspace{0pt}
    \centering
    \includegraphics[width=0.9\textwidth]{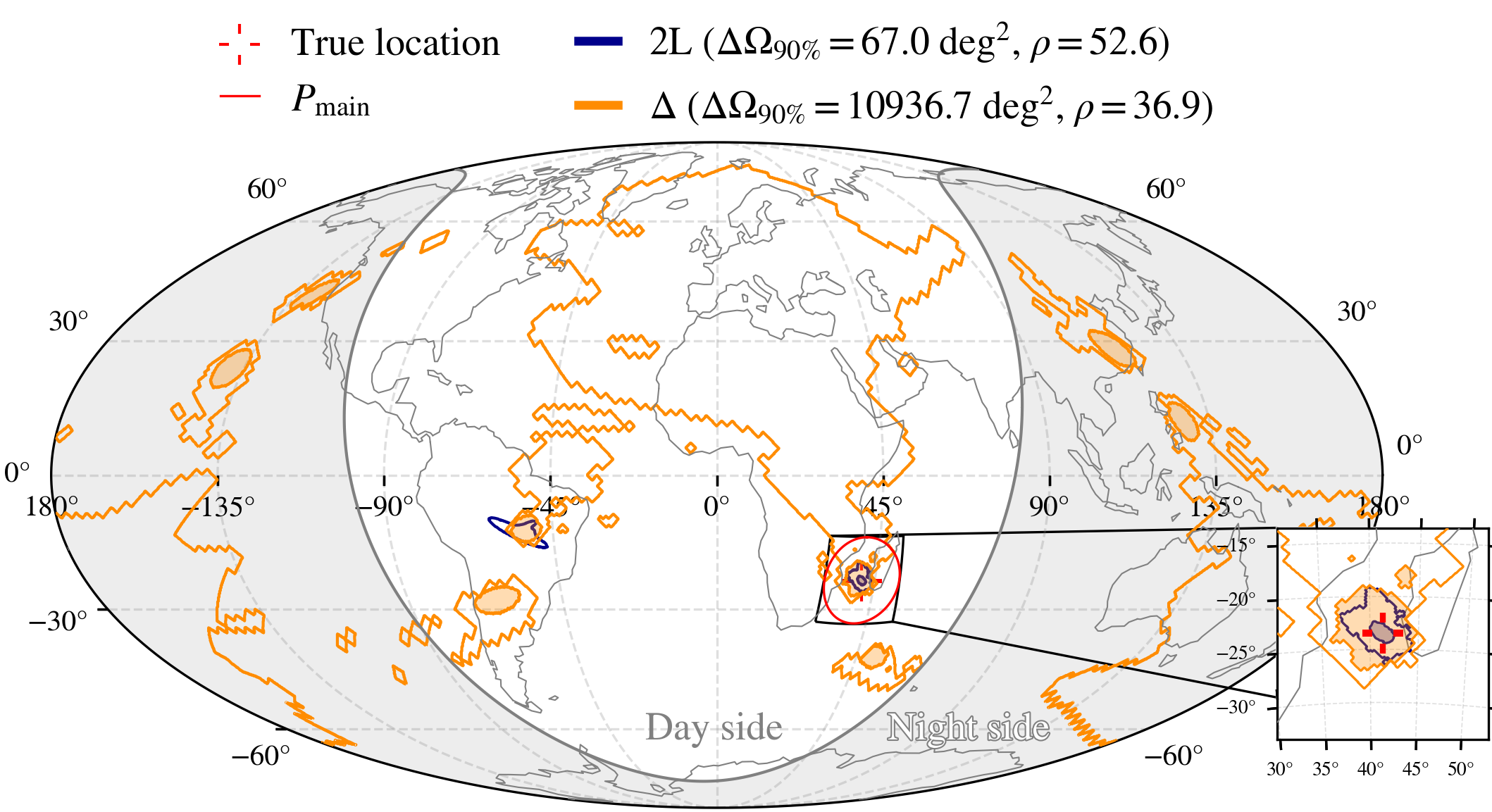}
\end{minipage}
\hspace{0.01\textwidth}
\begin{minipage}[t]{0.3\textwidth}
    \vspace{0pt}
    \centering
    \includegraphics[width=0.9\textwidth]{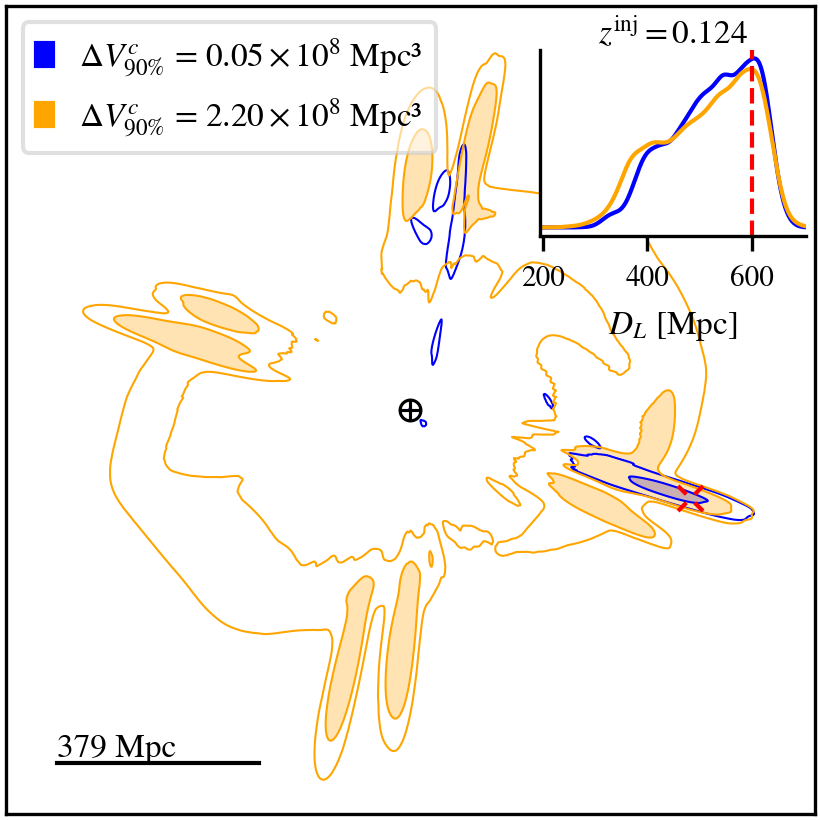}
\end{minipage}

\rowtitle{Low SNR, high-frequency sensitivity}
\begin{minipage}[t]{0.6\textwidth}
    \vspace{0pt}
    \centering
    \includegraphics[width=0.9\textwidth]{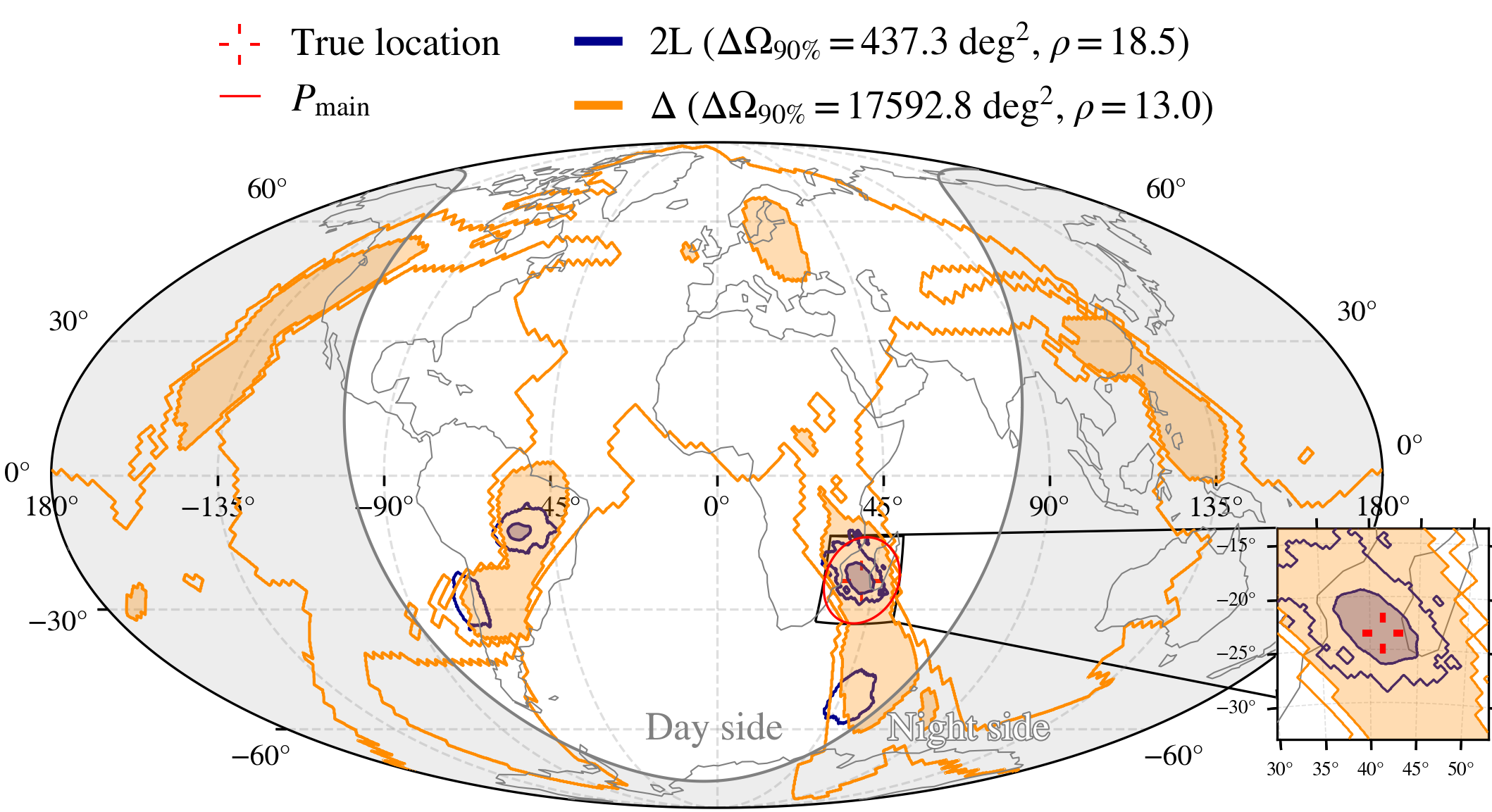}
\end{minipage}
\hspace{0.01\textwidth}
\begin{minipage}[t]{0.3\textwidth}
    \vspace{0pt}
    \centering
    \includegraphics[width=0.9\textwidth]{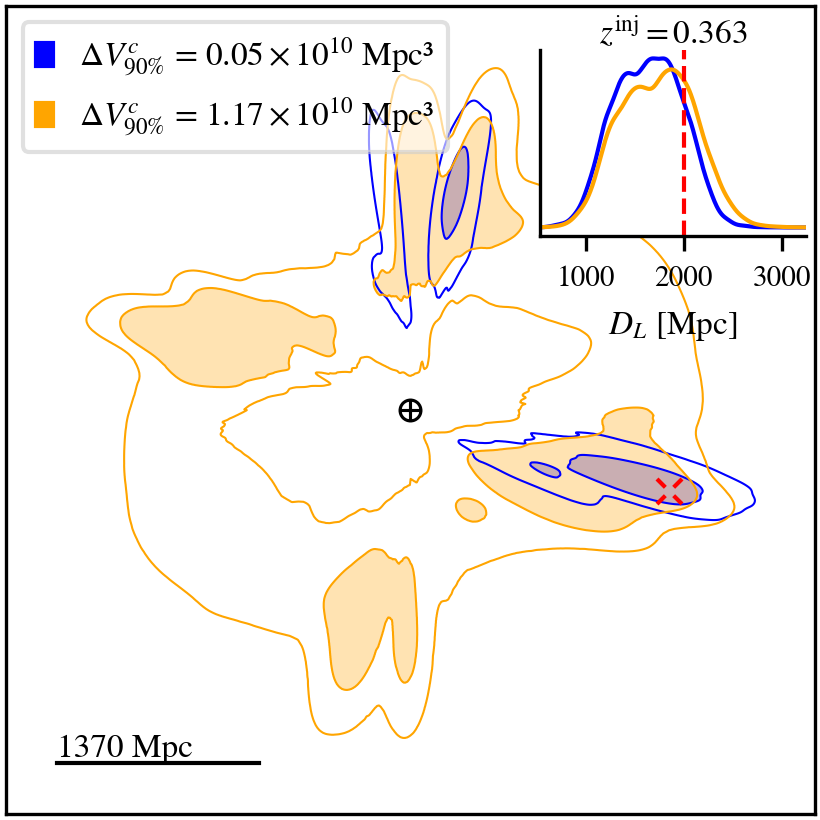}
\end{minipage}

\rowtitle{Low SNR, design sensitivity}
\begin{minipage}[t]{0.6\textwidth}
    \vspace{0pt}
    \centering
    \includegraphics[width=0.9\textwidth]{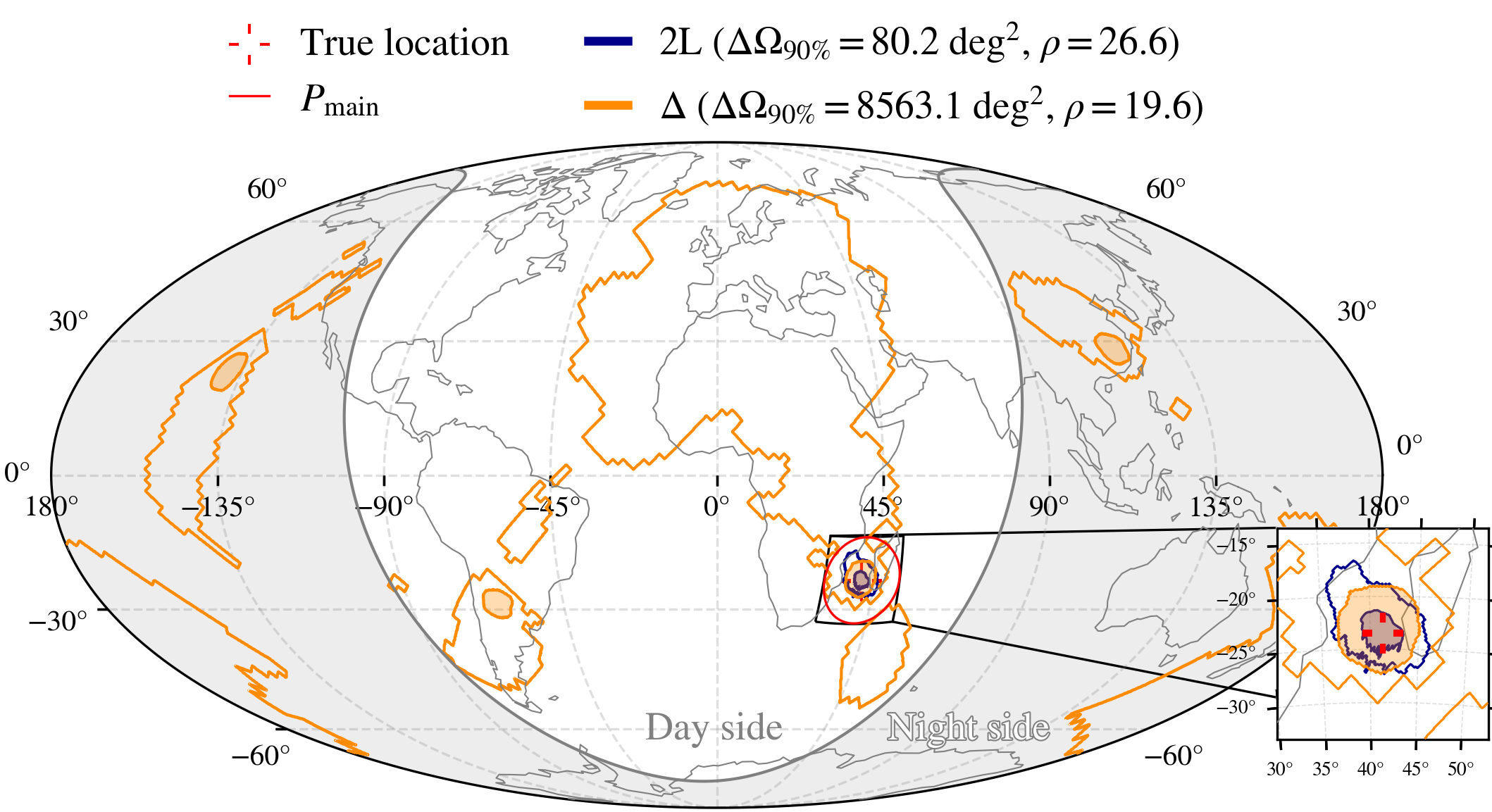}
\end{minipage}
\hspace{0.01\textwidth}
\begin{minipage}[t]{0.3\textwidth}
    \vspace{0pt}
    \centering
    \includegraphics[width=0.9\textwidth]{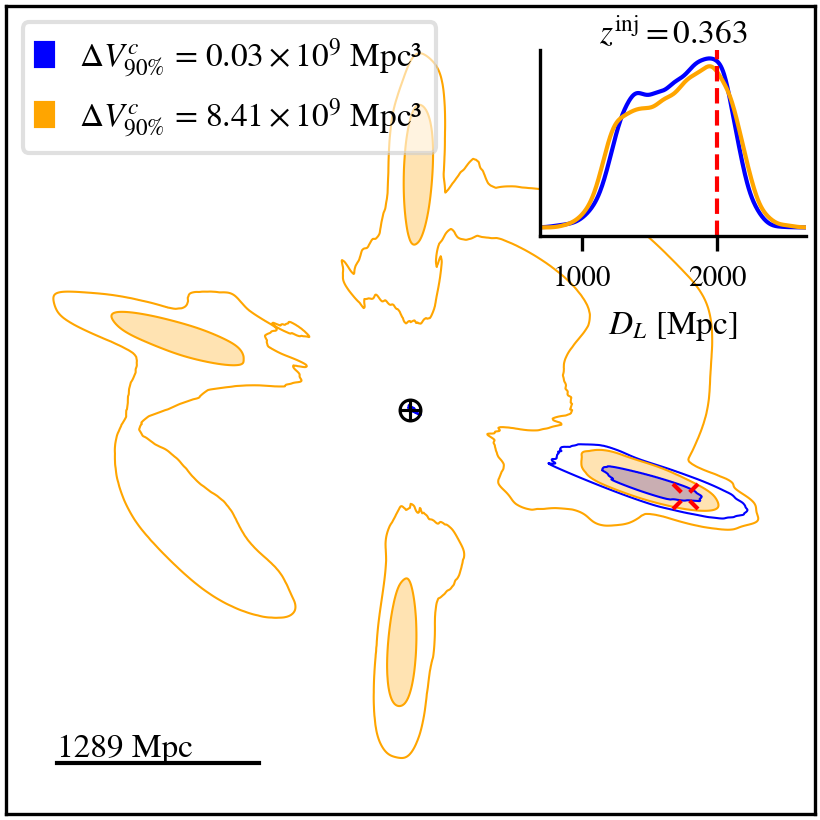}
\end{minipage}
\caption{Sky and volume localization of BNS signals, comparing the 2L (blue) and $\Delta$ (orange) configurations of ET. \textit{Top row:} moderate-SNR signal, HF sensitivity, i.e. with $f_\mathrm{min} = 6$~Hz and the HF instruments alone. \textit{Middle row:} low-SNR signal, HF sensitivity. \textit{Bottom row:} the same low-SNR signal as the middle row, but analyzed from $f_\mathrm{min} = 3$~Hz at the full design sensitivity (Appendix~\ref{sec:detector_appendix}). \textit{Left column:} $50\%$ and $90\%$ credible sky-localization contours, with the legend reporting the $90\%$ credible sky area $(\Delta\Omega_{90\%})$ and the optimal network SNR $(\rho$, see Appendix~\ref{sec:likelihood_appendix}) for each configuration. The shaded band marks the night side of the Earth at the time of merger (for GW170817, this was 2017 Aug 17, 12:41:04 UTC). The inset zooms onto the localization mode around the true position, marked by the red cross. The thin red circle marks the $300~\mathrm{deg}^2$ region around the true position used to define $P_{\mathrm{main}}$ (see Section~\ref{sec:pe}). \textit{Right column:} the 50\% and 90\% credible localization volumes, projected onto the equatorial plane as seen looking down the north pole, with the Earth at the center ($\oplus$). The corresponding 90\% credible comoving volume $(\Delta V^c_{90\%})$ is given in the legend. The inset shows the marginal luminosity distance posterior, with the injected value marked by the red dashed line. The top two rows are discussed in Section~\ref{sec:low_snr}, the bottom row in Section~\ref{sec:low_freq}.}
\label{fig:4panel}
\end{figure*}

\begin{figure}
    \centering
    \includegraphics[width=0.5\linewidth]{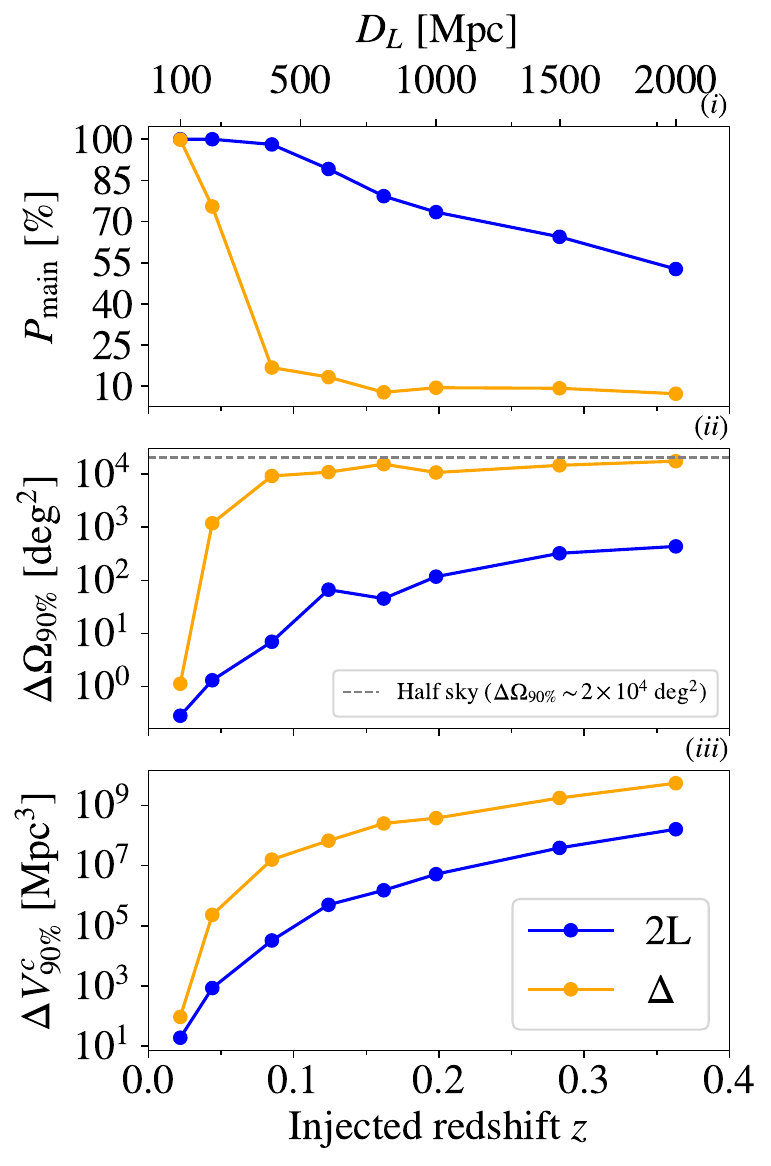}
    \caption{Localization performance of BNS signals as a function of injected redshift ($z$, bottom axis) and luminosity distance ($D_L$, top axis), at fixed sky localization, comparing the 2L (blue) and $\Delta$ (orange) configurations of ET. From top to bottom: $(i)$ probability mass contained in the mode centered on the injected value ($P_{\mathrm{main}}$, Section~\ref{sec:pe}) ; $(ii)$ the $90\%$ credible sky area ($\Delta\Omega_{90\%}$), where the horizontal dashed line marks half of the solid angle of a sphere; and $(iii)$ the $90\%$ credible comoving volume ($\Delta V^{c}_{90\%}$). See Section~\ref{sec:low_snr} for details.}
    \label{fig:redshift}
\end{figure}

\begin{figure*}
    \centering
    \includegraphics[width=0.8\linewidth]{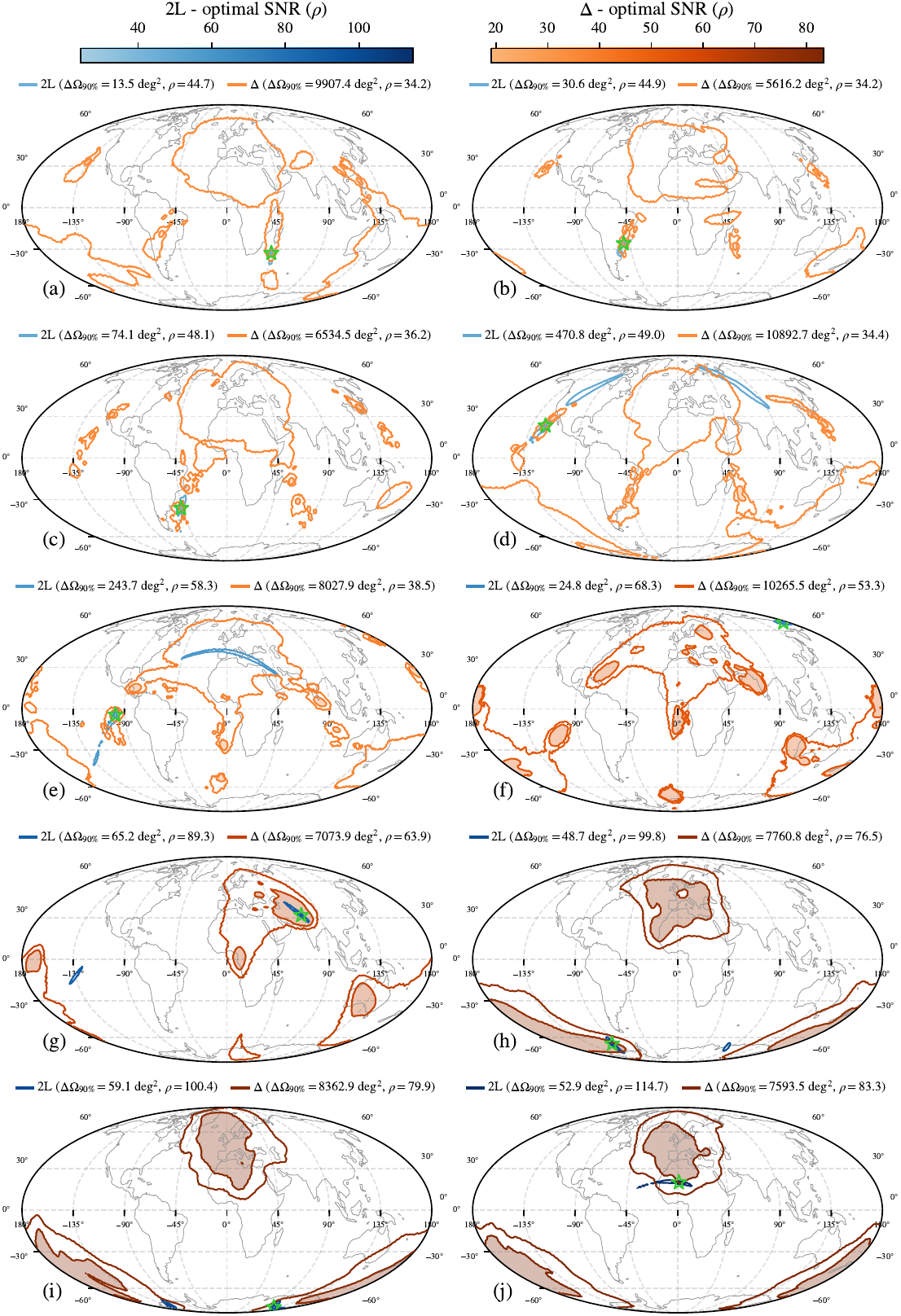}
\caption{BNS injections with fixed luminosity distance ($D_L^\mathrm{inj} = 600$~Mpc, $z^\mathrm{inj} = 0.124$) and inclination angle equal to 20° and varying true sky position, used to compare sky localization between the two ET geometries: the 2L configuration (blue) and the $\Delta$ configuration (orange). Panels show ten such events, ordered by increasing optimal SNR with the 2L configuration, with $50\%$ and $90\%$ credible regions and the true source position marked by a green star. The legend of each panel reports the $90\%$ credible sky area ($\Delta\Omega_{90\%}$) and the optimal SNR $(\rho)$ for each configuration. See Section~\ref{sec:low_snr} for details. Figure~\ref{fig:skypos_app} in Appendix~\ref{sec:low_snr_app} shows the same results but with inclination angle equal to $60$°.}
    \label{fig:skypos}
\end{figure*}

The upper panels of Figure~\ref{fig:4panel} show the results for two BNS injections, placed at luminosity distances of 600~Mpc and 2000~Mpc and recovered with both the $\Delta$ and 2L configurations. Both injections share the true sky location and merger time of GW170817 and inclination angle equal to 20°. They span the low- to moderate-SNR regime: the network SNRs are 36.9 and 13.0 for the $\Delta$ configuration and 52.6 and 18.5 for the 2L configuration, at the nearer and farther distance, respectively.

For low- and moderate-SNR BNS events, the $\Delta$ configuration localizes the source to eight distinct degenerate sky modes. We explain the origin of this feature in Appendix~\ref{sec:delta_skyloc} and discuss its implications in Section~\ref{sec:discussion}. These are distributed across the northern and southern hemispheres and over four different continents. The large angular separation between the modes places some on the night side and others on the day side of the Earth. This translates into a much larger sky-localization error for the $\Delta$ configuration, by a factor of $\sim 160$ at 600~Mpc and $\sim 40$ at 2000~Mpc relative to the 2L. 

The difference is larger for the louder, closer source: the 2L localization improves by a factor of $\sim 7$ going from the farthest to the nearest distance, whereas the $\Delta$ localization remains close to the half-sky scale and improves only marginally. The luminosity-distance uncertainties are comparable between the two configurations, so this degradation propagates into the comoving volume, for which the $\Delta$ configuration is a factor of $\sim 44$ and $\sim 23$ worse at the nearer and farther distance, respectively.

The 2L configuration also develops secondary modes at the lowest SNRs (mid panel of Figure~\ref{fig:4panel}), but they are far more compact in the celestial sphere. 
In Section~\ref{sec:low_freq} we show how this picture changes when the source is analyzed at design sensitivity.


Figure~\ref{fig:redshift} compares the sky-localization performance of the two configurations as a function of the injected redshift ($z^\mathrm{inj} \in [0.022, 0.363]$), at fixed injected sky position, again that of GW170817, and inclination angle equal to 20°. For the $\Delta$ configuration, the localization degrades rapidly as the source moves to higher redshift, with the main mode retaining around $10\%$ of the total probability mass at the highest redshift considered. The 2L configuration never degrades this far: its main mode retains at least $50\%$ of the probability mass even in the worst case. For the closest sources, at $z^\mathrm{inj} \lesssim 0.02$, corresponding to $\rho \gtrsim 300$ for the 2L configuration and $\rho \gtrsim 200$ for the $\Delta$, both recover essentially all of the probability mass in the main mode, with $P_\mathrm{main} \simeq 100\%$. We discuss this high-SNR regime further in Section~\ref{sec:high_snr}. 

Figure~\ref{fig:redshift} also shows that the sky areas and comoving volumes of the two configurations converge only in the high-SNR regime. For $\Delta$, the sky area increases abruptly by three orders of magnitude between the two nearest injections ($D_L = 100$ and $200\,$Mpc), from $\sim 1$ to $\sim 10^3\,\mathrm{deg}^2$, while the 2L area grows only by a factor of $\sim 4$. Over the rest of the redshift range, the $\Delta$ credible volumes exceed the 2L ones by more than an order of magnitude, and by more than two at $z \lesssim 0.1$.

Figure~\ref{fig:skypos} shows the results obtained by randomly varying the true sky position of the injected BNS system while keeping all other parameters fixed, with the inclination angle set to $20$° and the luminosity distance to 600~Mpc. The $\Delta$ configuration again performs poorly, displaying the distinctive eight-mode sky pattern for most sources, with $\Delta\Omega_{90\%}$ exceeding $10^4$~deg$^2$ in the worst of these cases (panel $(d)$ and $(f)$). The degradation is also severe when the true sky position approaches the local zenith at the detector site, i.e. where the SNR is largest, with $\Delta\Omega_{90\%} \sim 8000$~deg$^2$ in panels $(i)$ and $(j)$. The 2L configuration instead reconstructs the characteristic banana-shaped region familiar from BNS sources observed at comparatively modest SNR by the LIGO--Virgo--KAGRA detectors, such as GW170817 (Figure~3 of \textit{Abbott et al. 2017} \cite{LIGOScientific:2017vwq}), with $\Delta\Omega_{90\%}$ reaching at worst 470~deg$^2$, for the source shown in panel $(d)$. Appendix~\ref{sec:low_snr_app} repeats the same exercise for a wider inclination angle of $60$°, leading to qualitatively similar conclusions, although the reconstructed sky localizations are generally worse for both configurations because of the lower SNR.

Moreover, the 2L localizations in Figure~\ref{fig:skypos} are not a monotonic function of the SNR. The best-localized source, panel~$(a)$ at 13.5~deg$^2$, has the lowest optimal SNR of the ten, while panel~$(d)$ at $\rho = 49.0$ is localized to 470.8~deg$^2$, more than six times worse than panel~$(c)$ at a nearly identical $\rho = 48$. What matters for the 2L configuration over this range is not the loudness of the signal but where the source falls relative to the two interferometers: sources whose position is poorly constrained by the time-of-arrival difference between Sardinia and Lusatia produce extended banana-shaped regions regardless of their SNR, as in panels~$(d)$ and~$(e)$. The $\Delta$ configuration shows no comparable spread, since its localization is set by the eight-fold antenna-pattern degeneracy, which is a property of the geometry rather than of the source position.


\subsection{Low-SNR event at design sensitivity}
\label{sec:low_freq}

In this section, we analyze the source shown in the mid panel of Figure~\ref{fig:4panel}---the lowest-SNR source considered in this work---at the full design sensitivity of ET, again published in \textit{Dupletsa et al. (2026)} \cite{Dupletsa:2026jzs} and shown in Figure~\ref{fig:asds}. We adopt a minimum frequency of 3~Hz, for which the signal lasts 5~hours and 7~minutes in band, requiring a frequency grid of more than 75 million points to evaluate the full likelihood.

The optimal SNR is 26.6 (19.6) for the 2L ($\Delta$) configuration, compared to 18.5 (13.0) for the HF-sensitivity case in the mid panel of Figure~\ref{fig:4panel}. The sampling time is almost unchanged for the 2L configuration, whereas it took more than 8 days and 22 hours for the $\Delta$ configuration, instead of the 3~days and 6~hours required in the HF-only case.

The bottom panels of Figure~\ref{fig:4panel} show that the 2L configuration localizes the source to a single sky mode, corresponding to $\Delta\Omega_{90\%} = 80$~deg$^2$, whereas the $\Delta$ configuration retains the eight distinct sky modes, reaching $\Delta\Omega_{90\%} = 8500$~deg$^2$.
Compared to the HF-only configuration, the decrease in sky area is more than 5-fold for the 2L configuration, and only approximately 2-fold for the $\Delta$. In this example, the better localization capability of the 2L configuration compared to the $\Delta$ becomes even more significant when including the LF instrument than it was for the HF one only.

\subsection{High-SNR events}
\label{sec:high_snr}

\begin{figure*}
    \centering
    \includegraphics[width=0.6\linewidth]{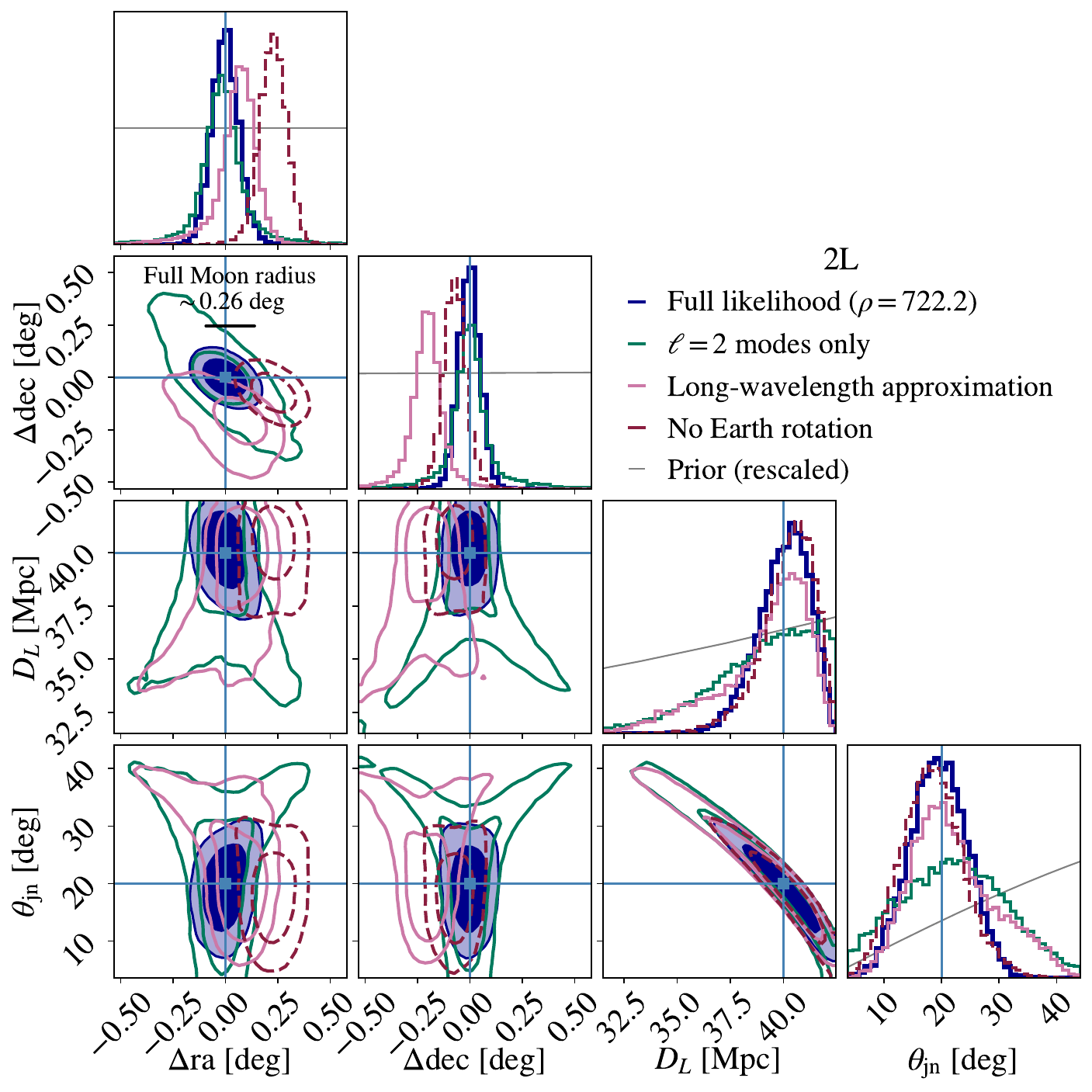}
    \vspace{0.5cm}
    \includegraphics[width=0.6\linewidth]{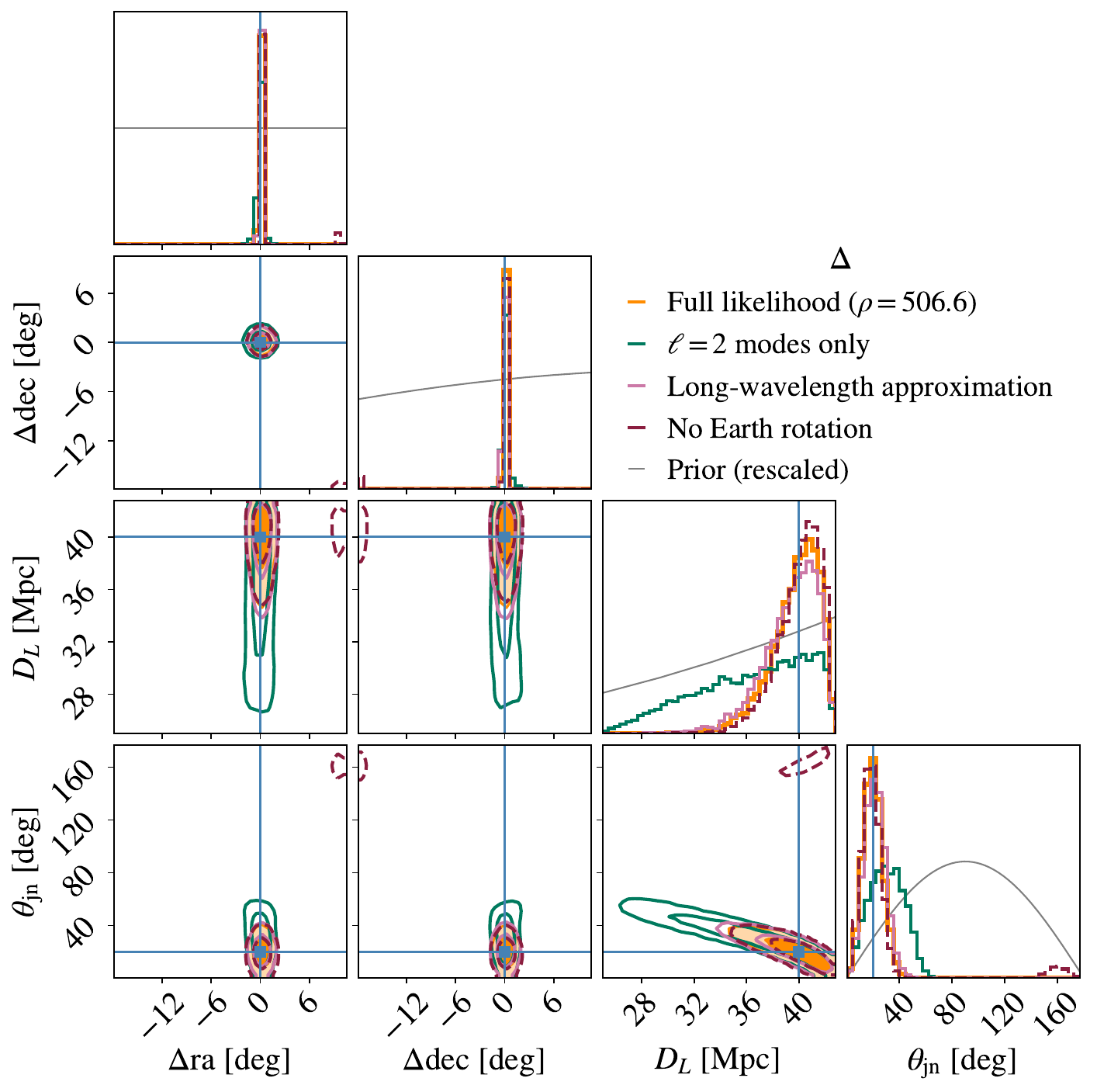}
    \caption{Marginalized posterior distributions (one- and two-dimensional) for the extrinsic parameters --- right ascension, declination, luminosity distance, and inclination angle --- for the 2L (upper panel) and $\Delta$ (lower panel) configurations of ET. We compare four analyses: full likelihood (blue for 2L, orange for $\Delta$), $\ell = 2$ modes only (teal), long-wavelength approximation (LWA, pink), and neglecting Earth rotation (dark red). Contours enclose the 68\% and 95\% credible regions, and vertical and horizontal lines mark the true injected values (Table~\ref{tab:injected}); right ascension and declination are shown as offsets from their injected values. For reference, the angular size of the full Moon is shown in the right ascension--declination panel of the 2L configuration only. The gray lines show the prior distributions, rescaled for visibility. See Section~\ref{sec:high_snr} for details.}
    \label{fig:high_snr}
\end{figure*}

Figure~\ref{fig:high_snr} shows a BNS system injected at a luminosity distance of 40~Mpc, i.e. with parameters close to those of GW170817, recovered with both ET configurations. The resulting network SNRs are 506 for the $\Delta$ configuration and 722 for the 2L configuration. At these SNRs both configurations localize the source with extremely high precision, down to the degree scale, comparable to the angular size of the full Moon.

Figure~\ref{fig:high_snr} also illustrates the impact of neglecting physical effects in the likelihood. Including HOMs significantly improves the localization and breaks the degeneracy between luminosity distance and inclination angle, demonstrating their constraining power \cite{LIGOScientific:2020stg}. 
The effect of the LWA and of Earth's rotation depends on the configuration. For 2L, adopting the LWA or neglecting Earth's rotation shifts the sky position by $\sim 0.15$--$0.25$~deg, in declination and right ascension respectively. This offset is small in absolute terms, but it is larger than the statistical uncertainty of the full-likelihood localization. For $\Delta$, the LWA has no visible effect, whereas neglecting Earth's rotation makes the sky localization multimodal: besides the mode at the true position, a secondary mode appears about 20~deg away, with mirrored inclination angle ($\theta_{\rm jn} \simeq 160$~deg). This is the reflection of the source through the plane of the three interferometers (Appendix~\ref{sec:delta_skyloc}). 


\section{Discussion}
\label{sec:discussion}




Our results show that both configurations can localize a high-SNR ($\rho \gtrsim 200$) BNS event with extremely high precision, down to the degree scale (see Figures~\ref{fig:redshift} and \ref{fig:high_snr}). The bulk of the BNS population detected by ET, however, will be observed at low-to-moderate SNR, typically below 100 \citep{Smith:2021bqc}. This expectation is reinforced by the absence of BNS detections in the fourth observing run (O4) of the LIGO-Virgo-KAGRA collaboration  \citep{LIGOScientific:2026wfs}, which points to a local merger rate at the lower end of previous estimates \citep{KAGRA:2021duu}, thereby reducing the expected number of nearby, louder events. Such a low rate is already in tension with the observed rate of short gamma-ray bursts \citep[GRBs;][]{DeSantis:2026wfw,Fishbach:2026vxv,Kunnumkai:2026jxw,Broekgaarden:2026bzj}, as it becomes challenging to attribute all detected short GRBs to BNS mergers.


Advanced optical survey instruments such as the Vera Rubin Observatory \citep{LSST:2008ijt} and WST \citep{Bisero:2025tkw}, expected to be still operational during the ET era, have a field of view of $\sim 10\,\mathrm{deg}^2$. Covering the sky map produced by the $\Delta$ configuration in Figure~\ref{fig:4panel} for low- and moderate-SNR events would therefore require several pointings and, since the modes are widely separated on the sky and in general not simultaneously observable from a single site, at least two different nights. Moreover, confirming a kilonova (KN) candidate requires repeated observations of each field in multiple filters to reject spurious sources and contaminating transients \citep{Andreoni:2021epw,Loffredo:2024gmx}, thereby substantially increasing the required number of pointings. Many candidates also require costly spectroscopic follow-up to be confirmed or ruled out as contaminants, and the number of contaminating transients increases with the surveyed area. 
Furthermore, the eight-modal sky-localization pattern characteristic of the $\Delta$ configuration makes detecting the early KN extremely challenging. Such a detection can provide unique information about the structure of the outermost ejecta and probe the presence of additional heating sources \citep{Banerjee:2020myd}. Similar arguments concerning the difficulty of a large number of sparse pointings apply at very high energies to instruments such as CTAO \citep{Banerjee:2022gkv}.

Covering all eight modes on a much shorter timescale is instead feasible at X-ray wavelengths from space, where the high-energy counterpart of the merger could be captured by wide-field monitors analogous to the Wide-field X-ray Telescope on board Einstein Probe, whose instantaneous field of view of $3600\,\mathrm{deg}^2$ \citep{Yuan:2022fpj} could enclose the entire sky pattern obtained with the $\Delta$ configuration in a few pointings. However, narrow-field X-ray observatories such as NewAthena, whose Wide Field Imager covers $0.44$ deg$^2$ \citep{Cruise:2024mgo}, would instead face the same tiling problem as ground-based facilities. 
Taken together, these considerations imply that the multi-messenger science case for the $\Delta$ configuration is, if not unfeasible, at best high-risk. The only exception is the prompt emission from GRBs, which can be detected by all-sky monitors operating in survey mode aboard high-energy satellites, although only for systems viewed on-axis \citep{Meegan:2009qu,Barthelmy:2005hs}.


Our modeling and analysis of BNS signals in ET account for the physical effects most relevant to sky localization: the higher-order modes of the emission and the frequency dependence of the antenna patterns, which captures both the finite size of the detector and the rotation of the Earth. Figure~\ref{fig:high_snr} shows that neglecting them is not harmless at high SNR. For 2L, the long-wavelength approximation or a non-rotating Earth shifts the sky position by more than its statistical uncertainty; for $\Delta$, a non-rotating Earth reintroduces a secondary sky mode at the mirrored inclination. The higher-order modes, in turn, improve the localization and break the degeneracy between luminosity distance and inclination angle.

Several elements are nonetheless absent from our treatment. We do not model non-stationary and non-Gaussian noise  \citep{Zackay:2019kkv,Kumar:2022tto}, nor the overlap of multiple signals \citep{Samajdar:2021egv} or glitches in the data stream \citep{Narola:2024qdh}, all of which will be unavoidable in ET and carry a substantial computational cost. A number of works have addressed these issues while retaining some level of  approximation \citep{Hu:2024lrj,Dax:2024mcn,Wouters:2024oxj,Iwaya:2026feu,Chatterjee:2024pbj,Guttman:2026fpc}.  We expect these effects to degrade the absolute localization accuracy of both configurations, while their impact on the relative comparison requires dedicated studies.


A second limitation is that we consider mainly
the HF instruments. The design-sensitivity runs of Section~\ref{sec:low_freq} suggest that including the LF band widens rather than narrows the gap between the two geometries: the 2L sky posterior collapses to a single mode, while the $\Delta$ retains its eight-fold multimodality, so that the improvement in sky area is more than fivefold for the former and only about twofold for the latter. 
This is also supported by a complementary study by \textit{Crescimbeni et al. (2026)}~\citep{Crescimbeni:2026ETLowFrequency}, who investigated the impact of site-dependent low-frequency noise on BNS early-warning and pre-merger localization for different ET configurations. For a GW170817-like source, they find that improved sensitivity at a few Hz substantially increases the early-warning time and improves sky localization through the longer signal duration and Earth-rotation modulation. Within the configurations they consider, the 2L network provides the tightest pre-merger localization, while the performance of the triangular layout varies significantly with the assumed site-dependent low-frequency sensitivity. Extending our analysis to the full design sensitivity is nonetheless computationally prohibitive at present: at
$f_\mathrm{min} = 3$~Hz the signal lasts more than five hours in band, which for the $\Delta$ configuration translated into almost nine days of sampling for a single low-SNR source, and would grow further with SNR. 

A third limitation is the restricted number of injections. A population-level study, which we defer to future work, would fold in the additional sources of uncertainty we have kept fixed: realistic merger rates, and hence a realistic distribution of SNRs, together with randomization over sky position, inclination, polarization and coalescence phase. Deep-learning based inference, such as {\sc Dingo} \citep{Dax:2024mcn}, offers a natural route to an analysis of that scale, and to the design-sensitivity comparison described above.


\section{Conclusions}
\label{sec:conclusion}

We present the first full Bayesian parameter estimation of BNS signals for ET that targets sky localization, comparing its two leading configurations: the $\Delta$ design at the EMR site and the 2L design with interferometers in Sardinia and Lusatia. Our
analysis relaxes the long-wavelength approximation, accounts for the rotation of the Earth, and includes the higher-order modes of gravitational wave emission, using the mode-by-mode relative-binning implementation of {\sc{Bilby-xg}} \citep{Baral:2025geo}. We consider a set of BNS injections with parameters close to those of GW170817, spanning optimal network SNRs from 12 to 720 and a range of true sky positions. 
We focus mainly on the ET observing scenario expected for the first years of operation, during which only the HF instruments will be online, as the LF instruments will become operational later due to the significant technical challenges involved.

For low- and moderate-SNR sources, which constitute the bulk of the BNS population detected by ET  \citep{Smith:2021bqc}, the $\Delta$ configuration reconstructs eight degenerate sky modes, spread over both hemispheres and over the day and night sides of the Earth (Figure~\ref{fig:4panel}). The mode containing the true position holds around $10\%$ of the posterior probability, and the $90\%$ credible sky areas reach $\sim 2\times 10^4~\mathrm{deg}^2$ (Figure~\ref{fig:redshift}). The 2L configuration instead recovers a single
dominant mode over the same SNR range, and retains at least $50\%$ of the posterior probability in the mode centered on the true position. Its $90\%$ credible sky areas are one to two orders of magnitude smaller than those of $\Delta$, with a corresponding improvement of up to about an order of magnitude in the credible comoving volume. The two geometries become
comparable only for the nearest and loudest events: at $\rho \gtrsim 200$ both localize the source to the degree scale (Figures~\ref{fig:redshift} and \ref{fig:high_snr}).




Fisher-matrix forecasts provided the general framework for comparing detector configurations through large-population studies, which already favored the 2L design over the triangular one for BNS localization \citep{Branchesi:2023mws,ET:2025xjr}. By properly accounting for the non-Gaussian features of the likelihood, our Bayesian analysis provides a more realistic forecast and shows that the gap is even larger.

These results have direct implications for the multimessenger science case of ET \citep{Branchesi:2023mws,ET:2025xjr}. The $\Delta$ configuration would require tiling multiple, widely separated sky regions that may not be simultaneously accessible from a single observing site, posing substantial challenges for the detection and identification of electromagnetic counterparts to BNS mergers. On the contrary, the significantly improved localization provided by the 2L configuration would greatly increase the chances of successfully identifying counterparts, allowing the ET to achieve its scientific objectives in the field of multi-messenger science with a significantly lower observational risk.

\section*{Data Availability}

Code and data will be made publicly available upon publication. 

\acknowledgments
We thank Eleonora Loffredo, Biswajit Banerjee, Samuele Ronchini, Sam Usman, Adriano Frattale Mascioli, Sebastiano Bernuzzi, Giulia Huez, Pratyusava Baral, Francesco Crescimbeni, Edoardo Milotti, and Simone Albanesi for insightful discussions. We also thank Anna Puecher and Patricia Schmidt for serving as internal reviewers for the Einstein Telescope collaboration. This manuscript has ET document number \href{https://apps.et-gw.eu/tds/?doc=ET-0748-26}{ET-0748A-26}. F.S. has been funded by the European Union – NextGenerationEU under the Italian Ministry of University and Research (MUR) "Decreto per l’assunzione di ricercatori internazionali post-dottorato PNRR" - Missione 4 "Istruzione e Ricerca" Componente 2 "Dalla Ricerca all’Impresa" del PNRR - Investimento 1.2 “Finanziamento di progetti presentati da giovani ricercatori” - CUP D13C25000700001. MB acknowledges the ACME project which has received funding from the European Union's Horizon Europe Research and Innovation program under Grant Agreement No.~101131928. 
J.T.~acknowledges funding from the Della Riccia Foundation under an Early Career Scientist Fellowship.

\appendix

\section{Detector configurations and noise spectral densities}
\label{sec:detector_appendix}

\begin{figure}
    \centering
\includegraphics[width=0.6\linewidth]{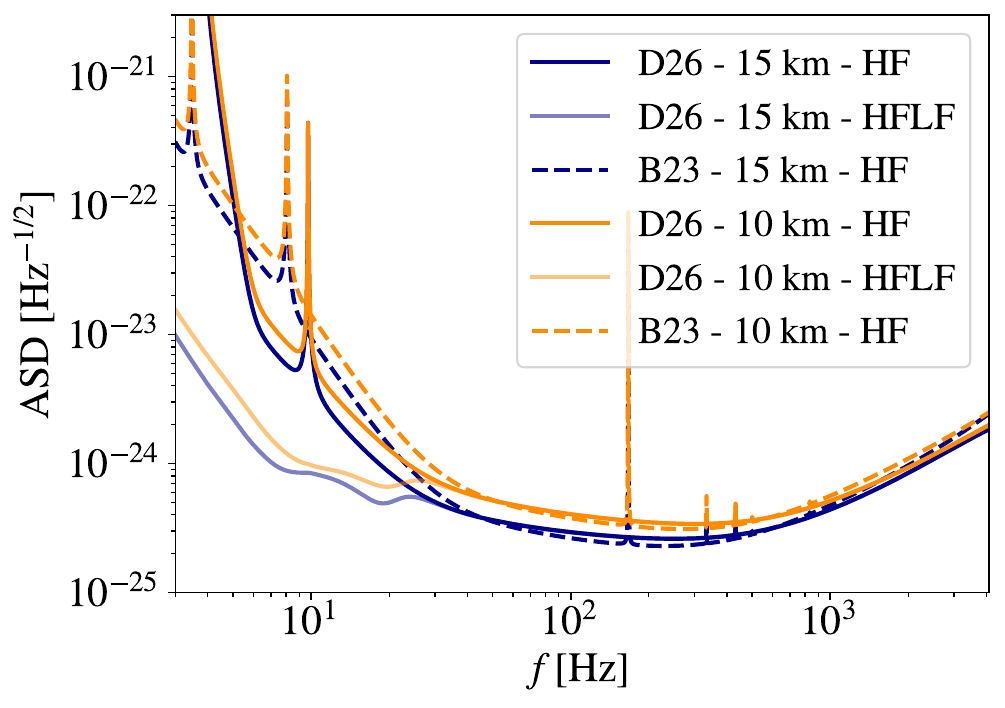}
    \caption{Amplitude spectral densities (ASDs) adopted in this work, from \textit{Dupletsa et al. (2026,~D26)} \cite{Dupletsa:2026jzs}, for the 2L configuration (15~km, blue) and the $\Delta$ configuration (10~km, orange). For each geometry, the D26 sensitivities are shown for the high-frequency instrument alone (HF, solid) and for the combined HF and LF instruments (HFLF, shaded). Dashed lines show the HF ASDs from \textit{Branchesi et al. (2023, B23)} \cite{Branchesi:2023mws} for comparison. See Section~\ref{sec:detectors} and Appendix~\ref{sec:detector_appendix} for details.}
    \label{fig:asds}
\end{figure}

\begin{table}[]
    \centering
    \begin{tabular}{cccc}
    \toprule
        Detector & Location & Coordinates & $x$-arm azimuth \\
    \midrule
        \multirow{3}{*}{$\Delta$} & EMR site 1 & 50.723°, 5.920° & 251°\\
                                  & EMR site 2 & 50.638°, 5.873° & 131° \\
                                  & EMR site 3 & 50.655°, 6.013° & 11° \\
    \midrule
        \multirow{2}{*}{2L} & Sardinia & 40.516°, 9.416° & $0^\circ$ \\
                            & Lusatia  & 51.275°, 14.100° & $45^\circ$ \\
    \bottomrule
    \end{tabular}
    \caption{Coordinates (latitude and longitude) and orientations ($x$-arm azimuth, measured counterclockwise from East) of the detector configurations considered in this work. See Section~\ref{sec:detectors} and Appendix~\ref{sec:detector_appendix} for details.}
    \label{tab:detectors}
\end{table}

Table~\ref{tab:detectors} summarizes the locations and orientations of the two detector configurations considered in this study, while Figure~\ref{fig:asds} shows the ASDs for each detector over the adopted frequency range, both for the HF instrument alone and at the full design sensitivity. The HF ASDs used in \textit{Branchesi et al. (2023)} \cite{Branchesi:2023mws} are less sensitive in 6~Hz $< f < 30$~Hz than those adopted in this work \citep{Dupletsa:2026jzs}.

\section{Frequency-dependent antenna patterns}
\label{sec:fdep}

\begin{figure*}
    \centering
    \includegraphics[width=0.7\linewidth]{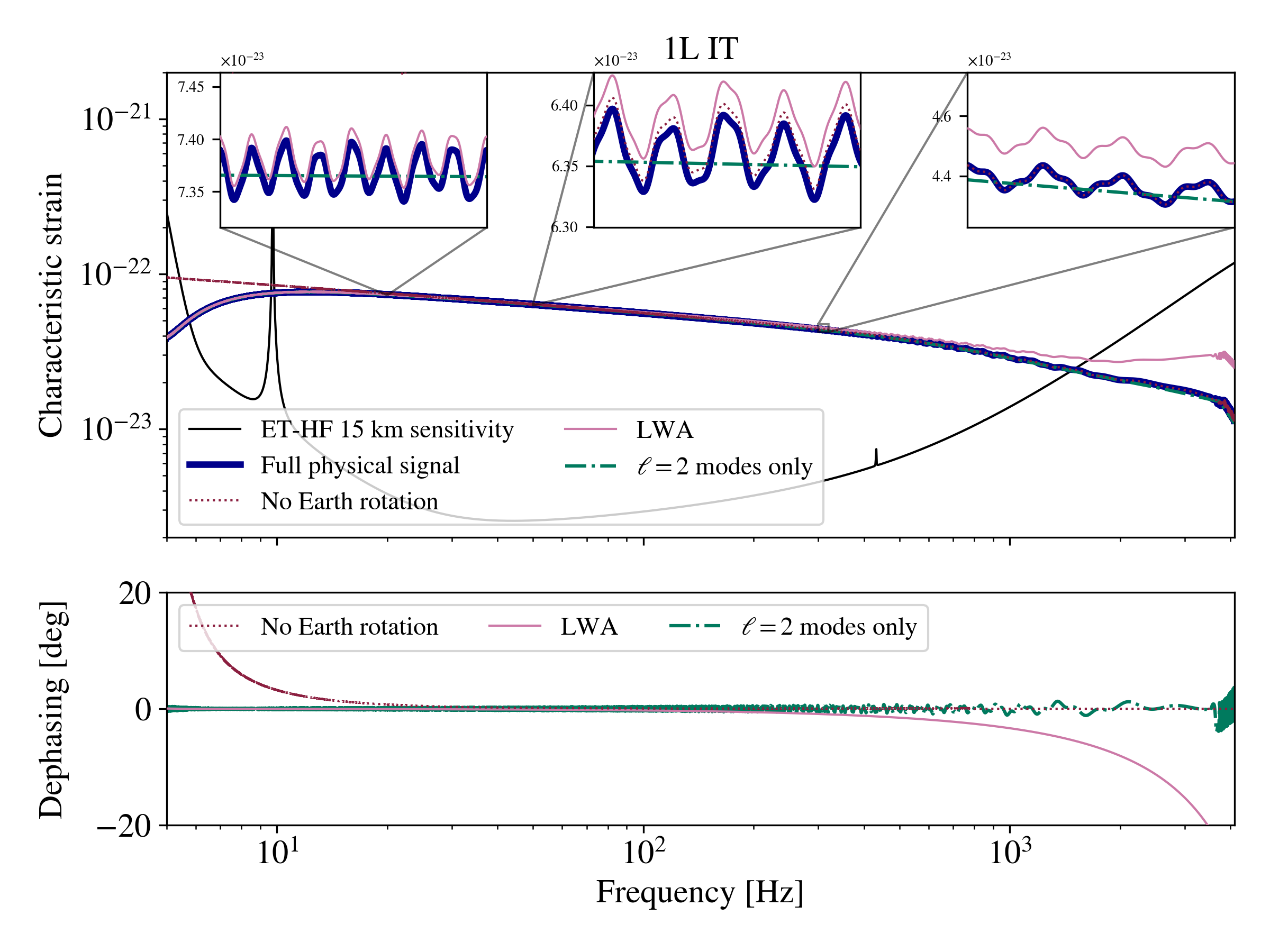}
    \caption{\textit{Top panel:} Characteristic strain of the BNS signal shown in the top panel of Figure~\ref{fig:4panel} observed with the L-shaped interferometer at the Italian site of the 2L configuration (1L IT). The thick dark blue curve shows the full detector response, including frequency-dependent antenna patterns. The other curves isolate individual effects: neglecting Earth rotation (dark red), adopting the long-wavelength approximation (LWA, pink), and retaining only the $\ell=2$ modes (green). The thin black curve shows the HF detector sensitivity (see also Appendix~\ref{sec:detector_appendix}). \textit{Bottom panel:} Phase difference of each approximation relative to the full response. See Appendix~\ref{sec:fdep} for details.}
    \label{fig:waveform}
\end{figure*}

The response of a GW interferometer is described by two antenna patterns, one for each polarization. These usually follow from the geometry of the interferometer under the long-wavelength approximation (LWA), which assumes that the interferometer is much smaller than the wavelength of the gravitational waves. The approximation holds as long as the GW frequency is well below the free spectral
range,
\begin{equation}
\label{eq:fsr}
    f_\mathrm{FSR} = \frac{c}{2L},
\end{equation}
where $c$ is the speed of light and $L$ the length of the interferometer arms. This condition is comfortably satisfied by current detectors, for which $f_\mathrm{FSR} = 37.5$~kHz with $L = 4$~km. The longer arms of next-generation instruments push it to much lower frequencies, down to $15.0$~kHz for the 10-km $\Delta$ and $10.0$~kHz for the 15-km 2L, so that the LWA eventually breaks down and the antenna patterns become frequency dependent~\citep{Whelan2007, Rakhmanov2005, Rakhmanov:2008is, Rakhmanov:2009zz}. This effect has been revisited more recently in the context of next-generation detectors~\citep{Essick:2017wyl}.

In the most general case, the detector response can be written as
\begin{equation}
\label{eq:response}
    \tilde{h}(f, \hat{n}) = \sum_{A \in \{+,\times\}}
    \tilde{F}_A(f,\hat{n},\psi)\,\tilde{h}_A(f),
\end{equation}
where $f$ is the signal frequency, $\hat{n}$ is the unit vector pointing to the GW source, $\psi$ the polarization angle, and $\tilde{F}_+(f,\hat{n},\psi)$ and $\tilde{F}_\times(f,\hat{n},\psi)$ are the frequency-dependent antenna patterns for the plus and cross polarizations, respectively. For a complete derivation we refer the reader to \textit{Rakhmanov et al. (2008)}~\cite{Rakhmanov:2008is}.

The line-of-sight vector $\hat{n}$ entering Equation~\ref{eq:response} changes with time, since the Earth rotates 15° per sidereal hour. In the frame co-rotating with the Earth, the source's declination stays fixed while its hour angle advances with time, and this motion introduces a further frequency dependence in the response.

The Earth's rotation enters the phase of the signal as well, through a Doppler modulation of the arrival time~\citep{Allen:2005fk}. The waveform
recorded at the detector is the geocentric waveform evaluated at the detector
arrival time, $t = t_\mathrm{geocent} - \Delta t$, where $\Delta t$ is the
light-travel time between the geocenter and the detector projected onto the line of sight \citep{Harry:2010fr,Singer:2015ema},
\begin{equation}
\label{eq:dt}
    \Delta t =
    \frac{\hat{n} \cdot \mathbf{d}}{c},
\end{equation}
and $\mathbf{d}$ points from the geocenter to the detector. For current detectors the signal is short enough that $\hat{n}$, and
hence $\Delta t$, is effectively constant over its duration. For the long-lived signals targeted by next-generation detectors this no longer
holds. As the Earth rotates, $\hat{n}$ moves across the sky while the source is in band, so $\Delta t$ varies over the signal \citep{Baral:2023xst, Baral:2025geo}. 

Figure~\ref{fig:waveform} illustrates the effects discussed above on the amplitude and phase of a BNS signal. The top panel shows the characteristic strain,
\begin{equation}
    h_c(f) = 2f\,|h(f)|,
\end{equation}
a dimensionless quantity representing the effective amplitude of a GW signal at a given frequency $f$ \citep{Moore:2014lga}, compared with the detector sensitivity $\sqrt{f\,S_n(f)}$, where $\sqrt{S_n(f)}$ is the amplitude spectral density (see Appendix~\ref{sec:detector_appendix}). 

Figure~\ref{fig:waveform} shows that neglecting Earth's rotation modifies the low-frequency portion of the signal, in both phase and amplitude: a BNS signal spends a long time in the ET band, up to several hours, so Earth rotates appreciably over the signal duration and modulates the detector response, an effect that is mostly visible at low frequencies. By contrast, the long-wavelength approximation (LWA) breaks down at high frequencies, where the finite light-travel time across the detector induces a frequency-dependent antenna response. Finally,
Figure~\ref{fig:waveform} compares a waveform containing only the fundamental $\ell = 2$ modes to one that also retains the higher-order modes, whose contribution likewise becomes relevant at high frequencies, particularly in the phase.





\section{Optimal signal-to-noise ratio, likelihood and relative binning}
\label{sec:likelihood_appendix}

\begin{figure*}
    \centering
    \includegraphics[width=0.75\linewidth]{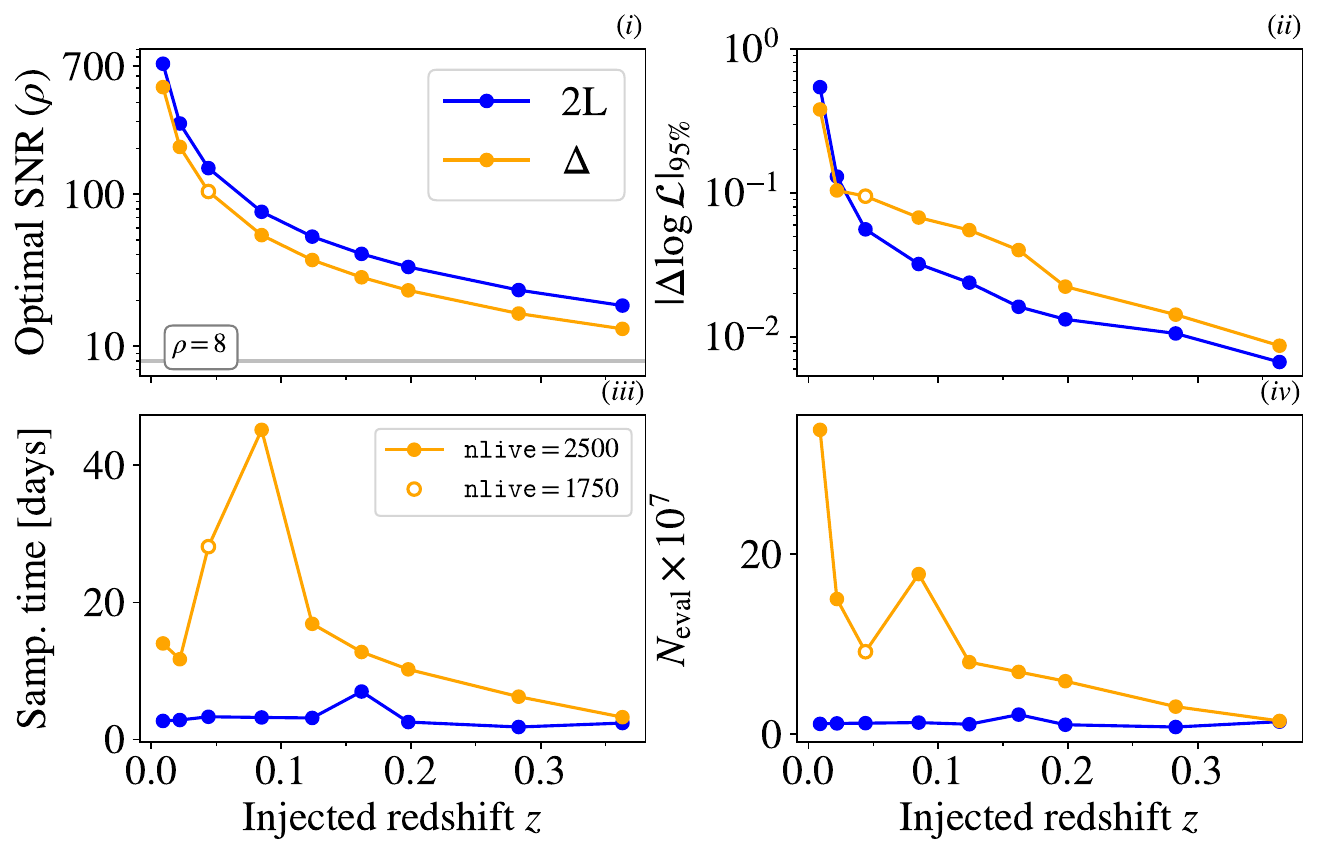}
    \caption{Inference performance of BNS signals as a function of injected redshift ($z$), comparing the 2L (blue) and $\Delta$ (orange) configurations of ET. From top left to bottom right: $(i)$ the optimal network SNR ($\rho$, Equation~\ref{eq:optimal_snr}), with the gray line marking the threshold $\rho = 8$; $(ii)$ the 95th percentile of the absolute differences between the exact log-likelihood and its relative-binning approximation; $(iii)$ the sampling time in days; and $(iv)$ the number of likelihood evaluations. See Appendix~\ref{sec:likelihood_appendix} for details.}
    \label{fig:redshift_app}
\end{figure*}

The noise-weighted inner product is defined as
\begin{equation}
\langle a \mid b \rangle = 4\,\mathrm{Re} \int_{f_{\rm min}}^{f_{\rm max}}
\frac{\tilde{a}^{*}(f)\,\tilde{b}(f)}{S_n(f)}\,\mathrm{d}f ,
\label{eq:inner_product}
\end{equation}
where $\mathrm{Re}$ denotes the real part, the asterisk the complex conjugate. 
This product is used to evaluate the optimal SNR,
\begin{equation}
\rho = \sqrt{\langle h(\theta^{\rm inj}) \mid h(\theta^{\rm inj}) \rangle} ,
\label{eq:optimal_snr}
\end{equation}
where $h(\theta^{\rm inj})$ is the GW signal evaluated at the injected parameters. Figure~\ref{fig:redshift_app} shows the range of optimal SNRs spanned by the sources analyzed in this work, from $\rho = 12$ to $\rho = 720$.

Assuming stationary, Gaussian noise, the likelihood is proportional to \citep{Finn:1992wt,Veitch:2014wba,Ashton:2018jfp}
\begin{equation}
\mathcal{L} (d | \theta) \propto \prod_i\exp\left(
\langle d_i \mid h_i(\theta) \rangle
- \frac{1}{2}\,\langle h_i(\theta) \mid h_i(\theta) \rangle
\right) \,,
\label{eq:likelihood_network}
\end{equation}
where the index $i$ labels the detectors in the network and $h_i(\theta)$ is the gravitational-wave signal projected onto detector $i$. In general, the data $d$ entering the likelihood comprise both a signal and a noise realization. In the present analysis, we perform zero-noise
injections, where for each detector we set
$d_i = h_i(\theta^{\mathrm{inj}})$.

Evaluating the scalar products in Equation~\ref{eq:likelihood_network} directly as a discrete sum over frequency bins requires approximately $2 \times 10^7$ points and about a minute of CPU time. We accelerate this using relative binning (RB), which exploits the fact that, for parameter values within the posterior's typical volume, the ratio between any two waveforms varies smoothly with frequency \citep{Zackay:2018qdy,Dai:2018dca,Cornish:2021lje} and can therefore be computed accurately on a sparse grid.

For systems with substantial contributions from HOMs---such as those with very high SNRs (of order hundreds) \citep[][]{Mills:2020thr,Kapadia:2020kss,Puecher:2022sfm,Pitte:2023ltw,Yi:2026ucv,Mascioli:2025cnw,Fairhurst:2023beb}, strongly asymmetric masses such as neutron star--black hole binaries (NSBHs) \citep[][]{Vitale:2018wlg,Zhang:2023ceh,Iacovelli:2026xjj}, or precessional effects \citep{Kalaghatgi:2019log,Singh:2023aqh}---the oscillatory behavior of the frequency-domain waveform (see also Figure~\ref{fig:waveform}) can degrade the accuracy of the standard piecewise-linear approximation. Although \textit{Krishna et al. (2023)} \cite{Krishna:2023bug} and \textit{Gupta (2024)} \cite{Gupta:2024bqn} showed that the standard RB technique still performs well for asymmetric systems, we adopt a more robust variant, referred to as mode-by-mode RB, proposed by \textit{Leslie et al. (2021)} \cite{Leslie:2021ssu} and subsequently by \textit{Narola et al. (2024)}
 \cite{Narola:2023men} specifically to handle the difficulties introduced by HOMs. The idea is to split the waveform into its spherical-harmonic modes
$(\ell, |m|)$, apply the RB approximation separately to each one, and then recombine them.



We construct frequency grids following \textit{Zackay et al. (2018)} \cite{Zackay:2018qdy}. In their scheme, the phase difference between any waveform in the posterior and the fiducial waveform is modeled as a sum of power laws in frequency, with post-Newtonian--motivated exponents and coefficient variations bounded by $2\pi\chi$, where we set $\chi = 1$. Bin edges are then placed such that this maximal phase difference varies by less than $\epsilon = 0.05$~rad within each bin. This procedure yields 626 bins between 6 and 4096~Hz, independently of the signal SNR, reducing the likelihood evaluation to 12 milliseconds.

To validate the robustness of the relative-binning approximation, we evaluate, for $n = 500$ posterior samples per injection, the absolute difference between the relative-binning and exact log-likelihoods, defined in Equation~\ref{eq:likelihood_network}. The upper right panel of Figure~\ref{fig:redshift_app} shows the upper 95th percentile of this distribution, which as expected increases with the signal SNR; even in the worst case, however, the error is still below the commonly adopted threshold $|\log \mathcal{L} - \log\mathcal{L}_\mathrm{RB}| < 1$ \citep{Leslie:2021ssu,Baral:2025geo}.


For all the sources analyzed in this work, we use the high-performance computing (HPC) system of the Laboratori Nazionali del Gran Sasso (LNGS), which comprises several nodes, each equipped with two 18-core Intel Xeon E5-2697 v4 (Broadwell) CPUs at 2.30~GHz and 128~GB of memory. We allocate one node per injection, setting \texttt{npool~=~32} to parallelize the sampling. We reserve the full node memory to evaluate the full likelihood over the $n = 500$ posterior samples, an operation that takes a few hours per injection.

The bottom left panel of Figure~\ref{fig:redshift_app} shows that the sampling time, measured excluding cluster downtime and maintenance, is consistently higher for the $\Delta$ configuration, with most sources requiring several days. This is driven by the complex likelihood surface of the $\Delta$ configuration, which in turn is due to the antenna pattern symmetries (see also Appendix~\ref{sec:delta_skyloc}). The {\sc nessai}
normalizing flow proposals become inefficient in this context, as evidenced by the systematically higher number of likelihood evaluations (bottom right panel of Figure~\ref{fig:redshift_app}). For this reason, while all other runs use
\texttt{nlive} = 2500 live points, we reduced this to 1750 for one $\Delta$
injections at $z^\mathrm{inj} = 0.044$, which would
otherwise have been prohibitively slow.

A smaller number of live points increases the risk that the sampler misses secondary modes, so that the resulting sky maps might underestimate the localization uncertainty for this $\Delta$ injection. A dedicated convergence test would require an independent sampler; neural posterior estimation, such as {\sc Dingo} \citep{Dax:2024mcn}, is well suited to this purpose, since it is trained to cover the full posterior mass and is therefore less prone to missing modes.



\section{Prior and injected parameters}
\label{sec:injection_appendix}

\begin{table}
    \centering
    \renewcommand{\cellalign}{tl}   
    \resizebox{\textwidth}{!}{%
\begin{tabular}{lccc}
\toprule
Parameter & Symbol/Units & Injected value(s) & Prior  \\
\midrule
Source-frame primary mass & $m_{1,s}/\mathrm{M}_\odot$ & 1.48 &  \\
Source-frame secondary mass & $m_{2,s}/\mathrm{M}_\odot$ & 1.28 &  \\
Mass ratio & $q$ & 0.8649 & $\mathcal{U}(0.5, 1)$ \\
Polarization angle & $\psi$/rad & 1.57 & $\mathcal{U}(0,\pi)$ \\
Phase at coalescence & $\phi_c$/rad & 1.57 & $\mathcal{U}(0,2\pi)$ \\
Geocentric time & $t_\mathrm{geocent}$/s & 1187008882.4 & $\mathcal{U}(t^\mathrm{inj}_\mathrm{geocent} \pm 0.1)$ \\
First aligned spin & $\chi_1$ & 0.0029 & Equation~\ref{eq:chi_prior} \\
Second aligned spin & $\chi_2$ & 0.0014 & Equation~\ref{eq:chi_prior} \\
\midrule
Inclination angle & $\theta_{\mathrm{jn}}$/rad & 
\makecell{%
    Figs.~\ref{fig:4panel},\ref{fig:redshift},\ref{fig:skypos},\ref{fig:high_snr},\ref{fig:waveform},\ref{fig:redshift_app},\ref{fig:2L_TEO},\ref{fig:Delta_TEO}: 0.3491 \\
    Fig.~\ref{fig:skypos_app}: 1.0472} 
    & $\mathcal{U}(\sin \theta_\mathrm{jn}),\, \theta_\mathrm{jn} \in [0,\pi]$ \\
    \addlinespace
Luminosity distance & $D_L$/Mpc &
    \makecell{%
      Fig.~\ref{fig:4panel}: 
      \{600, 2000\} \\
      Figs.~\ref{fig:redshift},\ref{fig:redshift_app}: 
      \{40, 100, 200, 400, 600, 800,\\ 1000, 1500, 2000\} \\
          Fig.~\ref{fig:high_snr}: 40\\
          Figs.~\ref{fig:skypos},\ref{fig:waveform},\ref{fig:2L_TEO},\ref{fig:Delta_TEO},\ref{fig:skypos_app}
      : 600 \\
 } &
    Equation~\ref{eq:dl_prior} \\
\addlinespace
Redshift & $z$ &
    \makecell{%
      Fig.~\ref{fig:4panel}: \{0.124, 0.363\}\\
      Figs.~\ref{fig:redshift},~\ref{fig:redshift_app}:
      \{0.009, 0.022, 0.044, 0.085, 0.124, 0.162,\\ 0.198, 0.283, 0.363\} \\
      Fig.~\ref{fig:high_snr}: 0.009\\
      Figs.~\ref{fig:skypos},\ref{fig:waveform},\ref{fig:2L_TEO},\ref{fig:Delta_TEO},\ref{fig:skypos_app}: 0.124 \\
      } &  \\
\addlinespace
Detector-frame chirp mass & $\mathcal{M}_d/\mathrm{M}_\odot$ &
    \makecell{%
      Fig.~\ref{fig:4panel}:
      \{1.35, 1.63\}\\
      Figs.~\ref{fig:redshift},~\ref{fig:redshift_app}:
      \{1.20, 1.22, 1.25, 1.30, 1.35, 1.39,\\ 1.43, 1.54, 1.63\} \\
      Fig.~\ref{fig:high_snr}: 1.20\\
      Figs.~\ref{fig:skypos},\ref{fig:waveform},\ref{fig:2L_TEO},\ref{fig:Delta_TEO},\ref{fig:skypos_app}: 1.35 \\
      } &
    $\mathcal{U}(\mathcal{M}_d^\mathrm{inj} \pm 0.001)$ \\
\addlinespace
Right ascension & $\alpha$/rad &
    \makecell{%
      Figs.~\ref{fig:4panel},\ref{fig:redshift},\ref{fig:high_snr},\ref{fig:waveform},\ref{fig:redshift_app},\ref{fig:2L_TEO},\ref{fig:Delta_TEO}
      : 3.4462 \\
      Figs.~\ref{fig:skypos},\ref{fig:skypos_app}: 
      \{3.4846, 1.8447, 1.9224, 0.5917,
        1.0133, \\ 5.7088, 3.9881, 0.9694, 
        4.9444, 2.7467\} \\
    } &
    $\mathcal{U}(0,2\pi)$ \\
\addlinespace
Declination & $\delta$/rad &
    \makecell{%
      Figs.~\ref{fig:4panel},\ref{fig:redshift},\ref{fig:high_snr},\ref{fig:waveform},\ref{fig:redshift_app},\ref{fig:2L_TEO},\ref{fig:Delta_TEO}
      : $-0.4081$ \\
      Figs.~\ref{fig:skypos},\ref{fig:skypos_app}:
      \{$-0.5770$, $-0.4525$, $-0.6376$, $0.4058$, \\
        $-0.0763$, $1.1779$, $0.5586$, $-1.1491$, 
        $-1.3990$, $0.3465$\} \\
    } &
    $\mathcal{U}(\cos\delta), \delta \in [-\pi/2, \pi/2]$ \\
\midrule
Tidal parameters & $\Lambda_{i = 1,2}$ & Figs.~\ref{fig:2L_TEO},\ref{fig:Delta_TEO}: 300& $\mathcal{U}(0,2500)$ \\
\bottomrule
\end{tabular}
}
    \caption{Injected parameter values and adopted priors for the BNS signals analyzed in this work. Here $\mathcal{U}(a, b)$ denotes a distribution uniform between $a$ and $b$, and $\mathcal{U}(x^\mathrm{inj} \pm \epsilon)$ is shorthand for $\mathcal{U}(x^\mathrm{inj}-\epsilon,\, x^\mathrm{inj}+\epsilon)$. See Appendix~\ref{sec:injection_appendix} for details. }
    \label{tab:injected}
\end{table}

\begin{figure}
    \centering
    \includegraphics[width=0.75\linewidth]{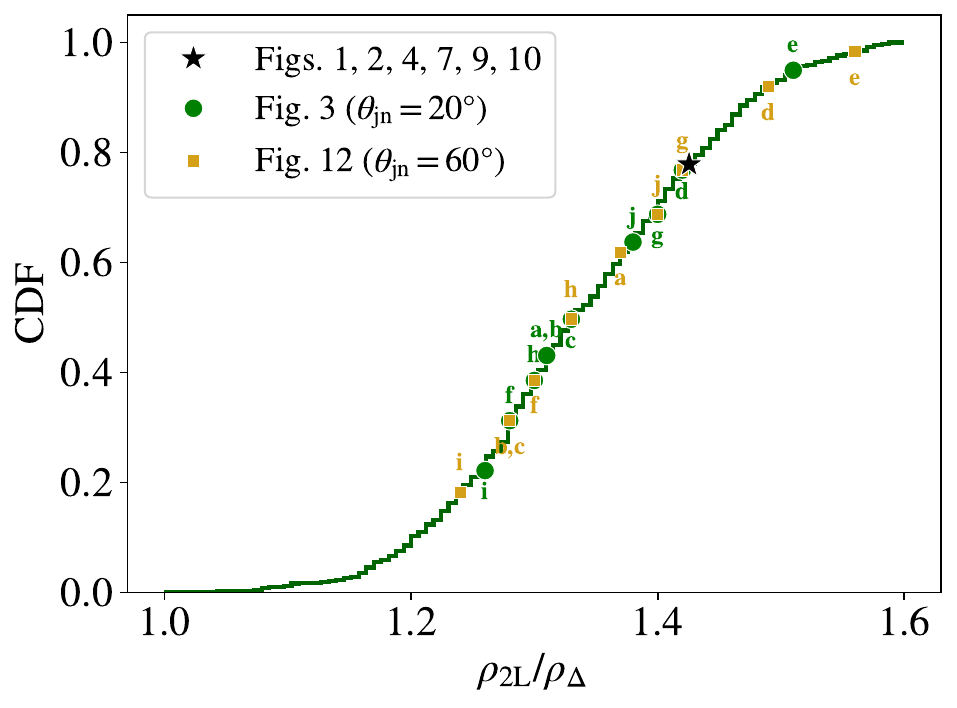}
    \caption{Cumulative distribution function (CDF) of the network optimal SNR ratio between the two ET configurations ($\rho_\mathrm{2L}/\rho_\Delta$), evaluated over 1000 BNS injections with inclination, sky position, phase, and polarization drawn randomly from the priors (Table~\ref{tab:injected}). Markers indicate the ratio for the injections analyzed in this work. The black star denotes the event of Figs.~\ref{fig:4panel}, \ref{fig:redshift}, \ref{fig:high_snr}, \ref{fig:redshift_app}, \ref{fig:2L_TEO}, and \ref{fig:Delta_TEO}, while green circles and gold squares denote the ten sky positions (labeled \textit{a}--\textit{j}) of Figures~\ref{fig:skypos} and \ref{fig:skypos_app}, for $\theta_\mathrm{jn} = 20^\circ$ and $60^\circ$, respectively. See Appendix~\ref{sec:injection_appendix} for details.}
    \label{fig:check_snr}
\end{figure}

The injected values of all the sources analyzed in this work, together with the chosen prior distributions $\pi(\theta)$, are listed in Table~\ref{tab:injected}. For the luminosity distance we adopt a prior that is uniform in comoving volume and source-frame time, assuming the \texttt{Planck18} \cite{Planck:2018vyg} cosmology,
\begin{equation}
    \label{eq:dl_prior}
    \pi(D_L) \propto \frac{\mathrm{d}V_c}{\mathrm{d}D_L}\frac{1}{1+z},
\end{equation}
with $D_L \in [15,\, 4000]$ Mpc. For the aligned spin component $\chi_i$, we construct a joint prior as
\begin{equation}
\label{eq:chi_prior}
    \pi(\chi_i) = \int_0^{0.05} \mathrm{d}a_i \int_{-1}^{1} \mathrm{d}\cos\theta_i\,\pi(a_i)\,\pi(\cos \theta_i)\, \delta(\chi_i - a_i\cos\theta_i),
\end{equation}
where $\pi(a_i) = \mathcal{U}(0, 0.05)$ and $\pi(\cos \theta_i) = \mathcal{U}(-1, 1)$ \citep{Lange:2018pyp,Ashton:2018jfp}.

The SNR of a source observed with the 2L configuration is on average higher than that of the same source observed with the 
$\Delta$ configuration, as a consequence of the reduced $60$° opening angle between the arms of the triangular design and their shorter length. The size of this advantage, however, varies from source to source, since it also depends on where the source falls in the antenna pattern of each configuration. The specific events analyzed in this work might therefore happen to be particularly favorable to 2L, which would bias the comparison in its favor.

To control for this, we evaluate the optimal SNR of 1000 sources, randomly sampling inclination, right ascension, declination, phase, and polarization from the priors (Table~\ref{tab:injected}), and compute the ratio between the SNRs of the two configurations. There is no need to vary the luminosity distance, as it cancels out in the ratio. Figure~\ref{fig:check_snr} shows that the injected values adopted in this work evenly span the distribution of SNR ratios, with only one source (panel \textit{(e)} of Figure~\ref{fig:skypos}) lying above the 90th percentile. The injected values adopted in the majority of the analyses performed in this work (Figs.~\ref{fig:4panel},~\ref{fig:redshift},~\ref{fig:high_snr},~\ref{fig:redshift_app},~\ref{fig:2L_TEO},~\ref{fig:Delta_TEO}) yield a ratio between optimal SNRs that lies below the 80th percentile.


\section{Impact of tides}
\label{sec:tides}

\begin{figure*}
    \centering
    \includegraphics[width=0.9\linewidth]{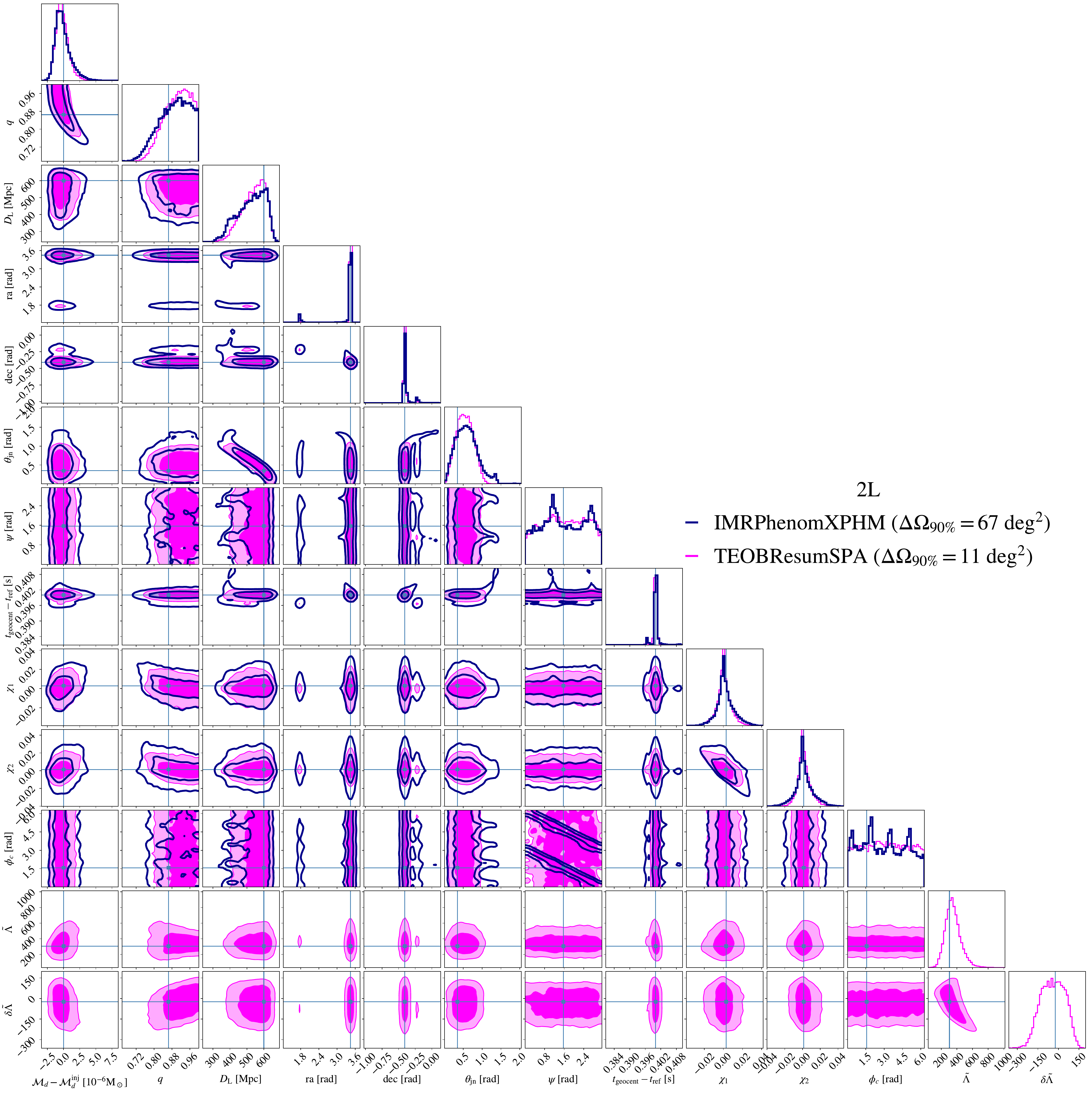}
    \caption{Marginalized one- and two-dimensional posterior distributions of all parameters for the event shown in the upper panel of Figure~\ref{fig:4panel}, as observed with the 2L configuration. Results obtained with \texttt{IMRPhenomXPHM} (dark blue) are compared with those obtained with \texttt{TEOBResumSPA} (magenta), which also includes tidal deformabilities ($\tilde{\Lambda}$ and $\delta\tilde{\Lambda}$). In both cases, the signal is injected and recovered with the same approximant. Vertical and horizontal lines mark the true injected values (see Table~\ref{tab:injected}), and contours represent the 68\% and 95\% credible regions. See Appendix~\ref{sec:tides} for further details.}
    \label{fig:2L_TEO}
\end{figure*}

\begin{figure*}
    \centering
    \includegraphics[width=0.9\linewidth]{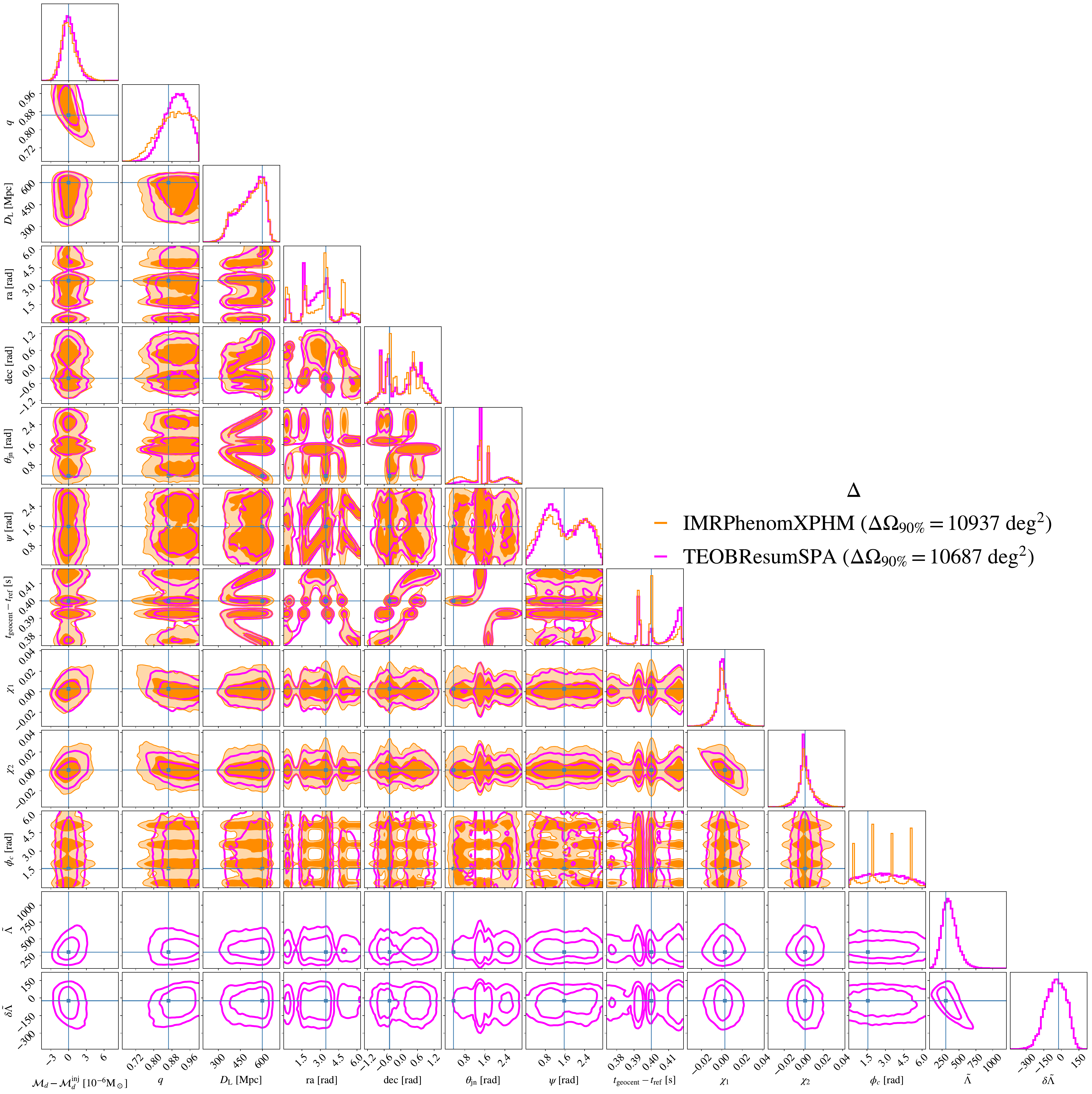}
    \caption{Same as Figure~\ref{fig:2L_TEO} but for the $\Delta$ configuration. See Appendix~\ref{sec:tides} for further details.}
    \label{fig:Delta_TEO}
\end{figure*}

The results in this work are obtained with the \texttt{IMRPhenomXPHM} waveform model, which describes the merger of binary black holes and is therefore not, in principle, representative of realistic BNS mergers. However, as we show here, the impact of the tidal interaction between the two neutron stars on the extrinsic parameters does not alter our qualitative conclusions.

To demonstrate this, we perform two injections with the fast, frequency-domain Effective-One-Body BNS model \texttt{TEOBResumSPA} \citep{Akcay:2018yyh,Gamba:2020ljo,Gamba:2023mww}, which provides an accurate treatment of this tidal effect.
We adopt the same setup and injection parameters as the source shown in the top panel of Figure~\ref{fig:4panel}, with the following changes: \emph{(i)} we use the \texttt{TEOBResumSPA} waveform model for both injection and recovery; \emph{(ii)} in addition to the parameters listed in Table~\ref{tab:injected}, we sample over the quadrupolar tidal parameters $\Lambda_{1,2}$ of the two stars, adopting a uniform prior in the range $[0, 2500]$ for both and injecting the values $\Lambda_{1}=\Lambda_{2}=300$, equivalent to $\tilde{\Lambda}=304$ and $\delta\tilde{\Lambda}=-23$; for simplicity, we assign the same value to both stars, which is compatible with current constraints for each of them, although it is not derived from any specific equation of state; \emph{(iii)} we include a slightly larger set of HOMs, adding the $(\ell,|m|)=(3,1)$ mode to those listed in Section~\ref{sec:pe}.
In this way, we account for all the modes that carry tidal corrections within the \texttt{TEOBResumSPA} model. As expected, inference with the complete waveform requires a substantially longer evaluation time per likelihood call: 95~ms compared to 12~ms with \texttt{IMRPhenomXPHM}.

We present the results of these injections in Figures~\ref{fig:2L_TEO} and \ref{fig:Delta_TEO}, for the 2L and $\Delta$ configurations respectively.
The marginal posterior distributions on the shared parameters are consistent between \texttt{IMRPhenomXPHM} and \texttt{TEOBResumSPA}.
The effective tidal parameter $\tilde{\Lambda}$ is constrained to ${341}_{-135}^{+186}$ (${353}_{-153}^{+217}$), while $\delta\tilde{\Lambda}$ is constrained to ${-39}_{-140}^{+131}$ (${-44}_{-147}^{+127}$) at $90\%$ credibility for the 2L ($\Delta$) configuration.
The sky localization is also consistent between the two approximants. For the $\Delta$ configuration the $90\%$ credible sky areas are comparable (10~937 and 10~687~deg$^2$ for \texttt{IMRPhenomXPHM} and \texttt{TEOBResumSPA}, respectively), while for the 2L configuration \texttt{TEOBResumSPA} yields a somewhat smaller area (67 and 11~deg$^2$)


Since it is constructed to model black-hole coalescences, the post-merger completion of \texttt{IMRPhenomXPHM} is the exponentially damped ringdown of a black hole, whereas the remnant of a neutron-star coalescence can exhibit long-lived post-merger emission. \texttt{TEOBResumSPA}, on the other hand, accounts only for the inspiral. Numerical-calibrated models for the post-merger emission of neutron-star binaries do exist \citep{Breschi:2022xnc,Breschi:2022ens}, but they add considerable complexity to the inference, and the comparatively quiet signals considered here do not justify their inclusion. \textit{Breschi et al. (2022)} \cite{Breschi:2022xnc} investigated the threshold for detecting this post-merger emission and found it to be around SNR~7 in the post-merger alone. Reaching this requires an overall signal SNR in excess of 1000 \citep{Dudi:2018jzn,Huez:2025gja}, which none of our injections attain.

\section{Interpreting $\Delta$ sky localization}
\label{sec:delta_skyloc}

\begin{figure}
    \centering
    \includegraphics[width=0.5\linewidth]{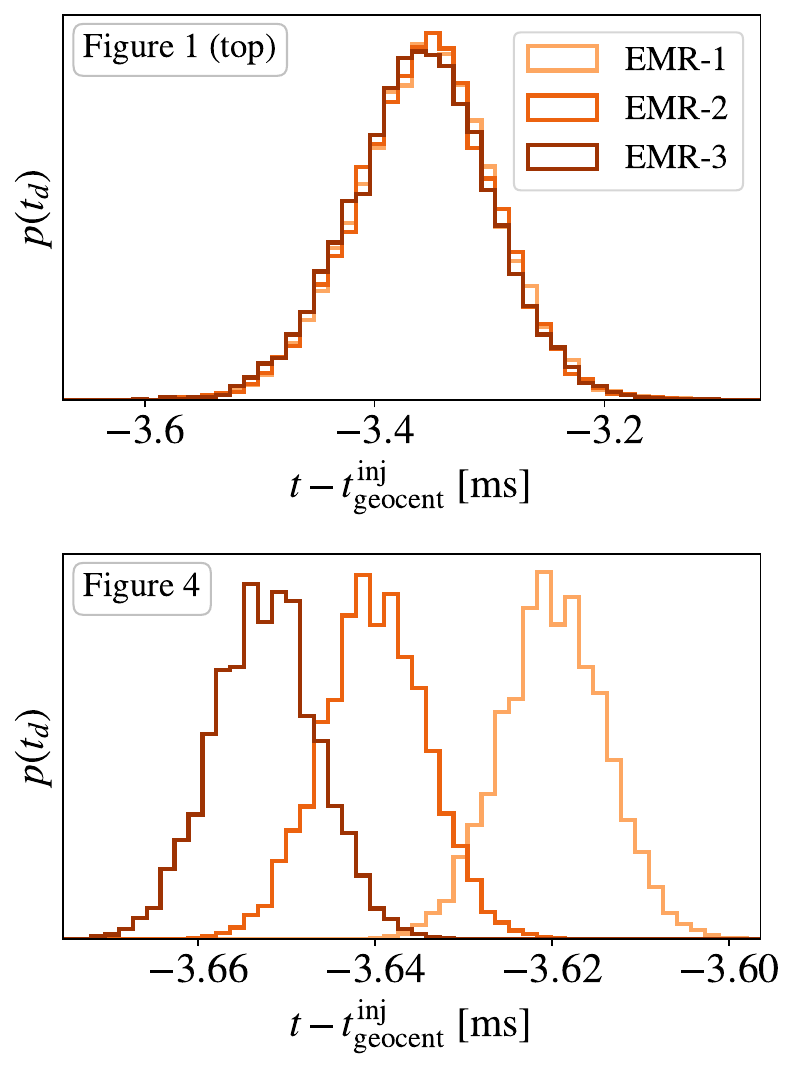}
    \caption{Arrival-time posteriors for two BNS injections recovered with the $\Delta$ configuration, at its three detectors (EMR-1, EMR-2, and EMR-3). The top panel corresponds to the moderate-SNR event in the upper panel of Figure~\ref{fig:4panel}, and the bottom panel to the high-SNR event in the bottom panel of Figure~\ref{fig:high_snr}. See Appendix~\ref{sec:delta_skyloc} for details.}
    \label{fig:time}
\end{figure}

The sky localization of short-lived binary black holes detected by ET in its triangular configuration exhibits eight distinct, separated modes. This feature is shared with the triangular space-based interferometer LISA, assuming frequency- and time-independent antenna patterns \citep{Marsat:2020rtl,Baibhav:2020tma,Santoliquido:2025lot}. The modes arise from the symmetries of the antenna pattern functions of the three detectors forming the equilateral triangle \citep{Regimbau:2012ir}: these produce eight degenerate sky positions in the posterior, evenly spaced in azimuthal angle by 90° and distributed symmetrically above and below the horizon. The degeneracy also imprints itself on the inclination angle, which consequently shows a bimodal distribution (see bottom panel of Figure~\ref{fig:high_snr}, Figure~\ref{fig:Delta_TEO}, and Figure~5 of \textit{Santoliquido et al. 2025} \cite{Santoliquido:2025lot}). Figure~\ref{fig:4panel} shows that these eight modes persist for long-lived BNS signals at low and moderate SNRs.

When the aforementioned symmetries are broken, through long-lived or high-frequency signals, the posterior support of the secondary modes is suppressed, and they may be effectively excluded \citep{Marsat:2020rtl}. There is a third symmetry-breaking avenue, however, only relevant for ET but not for LISA, due to the different typical ratio between signal wavelength and detector arm length. At sufficiently high SNRs, $\Delta$ can resolve the differences in time of arrival among its three interferometers, even though the $\sim 10$~km separation makes them extremely small (see also Figure~16 in \textit{Tissino et al. (2026)} \cite{Tissino:2026skt}). Figure~\ref{fig:time} shows the arrival-time posteriors at the three $\Delta$ detectors for the two representative BNS events. For the moderate-SNR injection of Figure~\ref{fig:4panel} (top panel), the three posteriors overlap entirely: the high-frequency content of the signal is insufficient to measure the sub-millisecond delays across the $\sim 10$~km baselines, leaving the degeneracies unbroken. For the high-SNR event of Figure~\ref{fig:high_snr}, which reaches $\rho \sim 500$ for $\Delta$, the three posteriors are instead clearly separated, showing that the delays are now resolved despite the short baselines and that the sky degeneracies are thereby lifted.

\section{Wider inclination angle at random true locations}
\label{sec:low_snr_app}

\begin{figure*}
    \centering
    \includegraphics[width=0.8\linewidth]{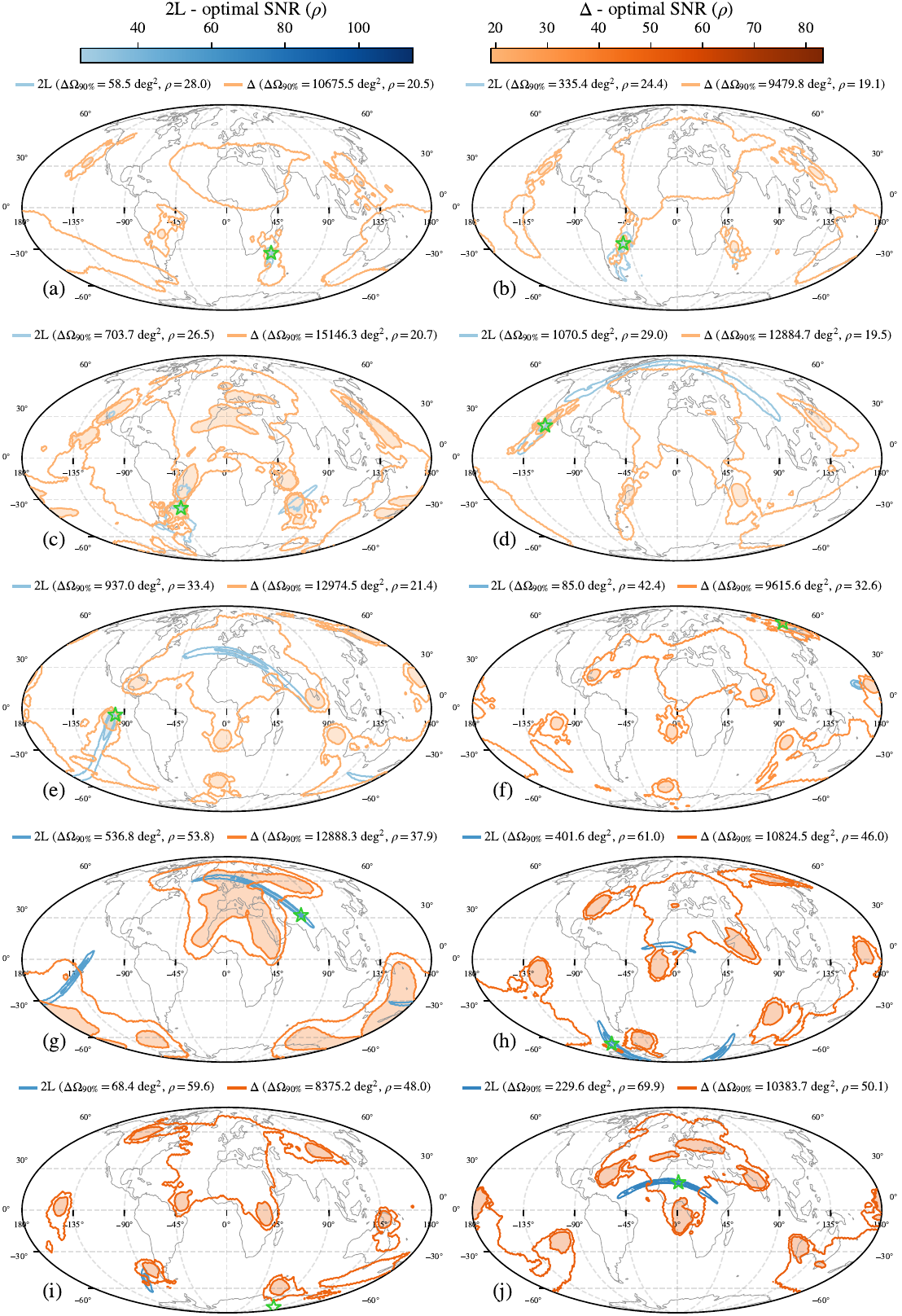}
\caption{BNS injections with fixed luminosity distance ($D_L^\mathrm{inj} = 600$~Mpc, $z^\mathrm{inj} = 0.124$) and inclination angle equal to $60$° and varying true sky position, used to compare sky localization between the two ET geometries: the 2L configuration (blue) and the $\Delta$ configuration (orange). Panels show ten such events in the same order of Figure~\ref{fig:skypos}, with $50\%$ and $90\%$ credible regions and the true source position marked by a green star. The legend of each panel reports the $90\%$ credible sky area ($\Delta\Omega_{90\%}$) and the optimal SNR $(\rho)$ for each configuration. See Appendix~\ref{sec:low_snr_app} for details. 
}
    \label{fig:skypos_app}
\end{figure*}

Figure~\ref{fig:skypos_app} shows the same sources analyzed in Figure~\ref{fig:skypos}, but with the injected inclination angle set to a wider value of $60$° instead of $20$°. As in Figure~\ref{fig:skypos}, all sources are placed at 600~Mpc ($z^{\rm inj} = 0.124$). The optimal SNRs drop by roughly a third for both configurations, and the sky localizations degrade accordingly, but they do not degrade equally. The 2L configuration is the more strongly affected: its $\Delta\Omega_{90\%}$ spans $\sim 13$–$470$~deg$^2$ at $20$°, with eight of the ten sources localized to better than $100$~deg$^2$, whereas at $60$° it spans $\sim 60$–$1100$~deg$^2$ and only three sources remain below that value. The best and worst cases worsen by factors of about four and two, respectively. For the $\Delta$ configuration the whole range shifts upward more modestly, from $\sim 5600$–$10900$~deg$^2$ to $\sim 8400$–$15100$~deg$^2$, but the number of sources whose localization exceeds $10^4$~deg$^2$ rises from two to seven out of ten. This asymmetry reflects the different origin of the two uncertainties: the $\Delta$ localization is limited by the eight-fold sky degeneracy rather than by the SNR, and is therefore already close to saturation at $20$°, whereas the 2L localization scales with the SNR and follows the inclination directly. Even at this less favorable inclination, the two configurations remain separated by one to two orders of magnitude in sky area.

\bibliographystyle{JHEP}
\bibliography{refs}

\end{document}